\documentclass[journal, 10pt]{IEEEtran}

\AtBeginDocument{
  \providecommand\BibTeX{{
    \normalfont B\kern-0.5em{\scshape i\kern-0.25em b}\kern-0.8em\TeX}}}

\usepackage{soul,xcolor}
\setstcolor{red}
\usepackage{graphicx}
\usepackage{float}
\usepackage{amsmath}
\usepackage{xcolor}
\usepackage{subcaption}
\usepackage[export]{adjustbox}
\usepackage{siunitx}
\usepackage{colortbl}
\usepackage{algorithmic}
\usepackage[thinlines]{easytable}
\usepackage{siunitx}
\usepackage[normalem]{ulem}
\usepackage{amsfonts}
\usepackage{multirow}
\usepackage{pgfplots}
\usepackage{url}
\usepackage{ctable} 
\usepackage{comment}
\usepackage[utf8]{inputenc}
\usepackage{gensymb}
\usetikzlibrary{decorations,shapes,arrows,positioning}
\usepackage{pifont}

\usepackage{color, colortbl}
\definecolor{lightgray}{RGB}{211,211,211}
\definecolor{lightcyan}{RGB}{224,255,255}

\newcommand{\cmark}{\ding{51}}%
\newcommand{\xmark}{\ding{55}}%

\newcommand{\crd}{\mathtt{C}}
\newcommand{\img}{\mathtt{I}}
\newcommand{\ldr}{\mathtt{L}}
  
\newcommand{\AGF}{\mathtt{AGF}}
\newcommand{\IRF}{\mathtt{IRF}}
\newcommand{\DF}{\mathtt{DF}}

\newcommand{\scores}{\mathbf{s}}

\begin{document}

\title{Going Beyond RF: How AI-enabled Multimodal Beamforming will Shape the NextG Standard}

\author{
		\IEEEauthorblockN{
        Debashri Roy\IEEEauthorrefmark{1},
        Batool Salehi\IEEEauthorrefmark{1},
        Stella Banou\IEEEauthorrefmark{1},
        Subhramoy Mohanti\IEEEauthorrefmark{1},
        Guillem Reus-Muns\IEEEauthorrefmark{1},
        Mauro Belgiovine\IEEEauthorrefmark{1},
        Prashant Ganesh\IEEEauthorrefmark{2},
        Carlos Bocanegra\IEEEauthorrefmark{1},
        Chris Dick\IEEEauthorrefmark{3}, and
        Kaushik Chowdhury\IEEEauthorrefmark{1}
              }
		    \IEEEauthorblockA{
		        \resizebox{0.9\textwidth}{!}{\begin{tabular}{ccc}
		            \begin{tabular}{@{}c@{}}
		                \IEEEauthorrefmark{1}
		   Electrical and Computer Engineering\\Northeastern University\\ Boston, MA 02115\\
		    \{droy, bsalehihikouei, sbanou, smohanti, \\greusmuns, mbelgiovine, cbocanegra, krc\}@ece.neu.edu
		            \end{tabular} & 
                    \begin{tabular}{@{}c@{}}
		                \IEEEauthorrefmark{2}
		    Mechanical and Aerospace Engineering\\University of Florida REEF\\  Shalimar, FL 32579\\ {prashant.ganesh@ufl.edu}
		            \end{tabular} & 
                    \begin{tabular}{@{}c@{}}
		                \IEEEauthorrefmark{2}
		    Nvidia Inc.\\ {cdick@nvidia.com}
		            \end{tabular}
		        \end{tabular}
		    }}\\\vspace*{-1cm}
		}

\maketitle

\begin{abstract}
Incorporating artificial intelligence and machine learning (AI/ML) methods within the 5G wireless standard promises autonomous network behavior and ultra-low-latency reconfiguration. However, the effort so far has purely focused on learning from radio frequency (RF) signals. Future standards and next-generation (nextG) networks beyond 5G will have two significant evolutions over the state-of-the-art 5G implementations: (i) massive number of antenna elements, scaling up to hundreds-to-thousands in number, and (ii) inclusion of AI/ML in the critical path of the network reconfiguration process that can access sensor feeds from a variety of RF and non-RF sources. While the former allows unprecedented flexibility in `beamforming', where signals combine constructively at a target receiver, the latter enables the network with enhanced situation awareness not captured by a single and isolated data modality. This survey presents a thorough analysis of the different approaches used for beamforming today, focusing on mmWave bands, and then proceeds to make a compelling case for considering non-RF sensor data from multiple modalities, such as LiDAR, Radar, GPS for increasing beamforming directional accuracy and reducing processing time. This so called idea of {\em multimodal beamforming} will require deep learning based fusion techniques, which will serve to augment the current RF-only and classical signal processing methods that do not scale well for massive antenna arrays. The survey describes relevant deep learning architectures for multimodal beamforming, identifies computational challenges and the role of edge computing in this process, dataset generation tools, and finally, lists open challenges that the community should tackle to realize this transformative vision of the future of beamforming.

\end{abstract}


\section{Introduction}





Today's ultra-connected world is demanding high bandwidths, ultra-low latency, and autonomous network reconfiguration to accommodate new applications, heterogeneous devices and shared spectrum use. 
The number of users is also increasing at unprecedented levels, with predictions of the number of networked devices exceeding $3x$ the global population by 2023 ~\cite{Cisco_2020}. To serve bandwidth-hungry application needs, the expected maximum 5G data rate is now revised to be $13x$ faster in 2023, a significant revision from earlier estimations made only a few years ago in  2018~\cite{Cisco_2020}. 
Many exciting applications will leverage such high capacity wireless networks, such as  relaying high-resolution three dimensional (3D) graphical content,  VR/AR streams~\cite{Haerick_2015}, vehicle-to-everything (V2X) links leading towards autonmous cars~\cite{Huawei_2014, Hossain_2014}, among others. 

\begin{figure}[!t]
\centering
\includegraphics[width=0.48\textwidth]{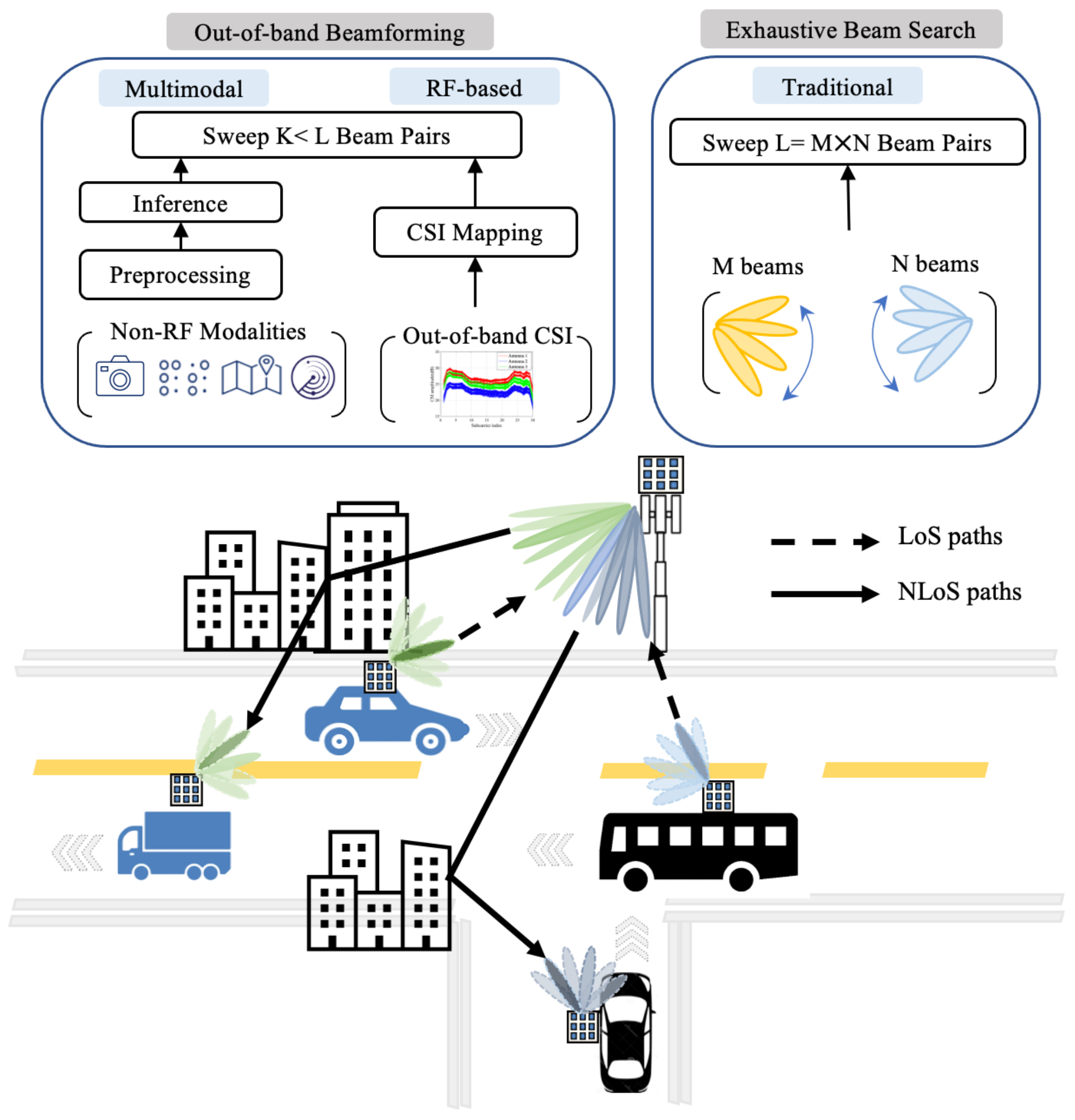}
\caption{\small An overview of different approaches for beamforming in an example scenario involving a mmWave vehicular network. The goal of beamforming between the roadside base station (BS) and the vehicle is categorized into three types: (a) {\em traditional exhaustive beam search} that sweeps through all possible beam combination between the receiver and transmitter, (b)  {\em RF-based out-of-band beamforming} that uses channel state information (CSI) measurements from lower frequencies to restrict mmWave beam search space, (c){\em multimodal beamforming} that uses non-RF sensor modalities (image, LiDAR, GPS, radar) to predict the best possible beams from the situational information.}
\label{fig_overview}
\vspace{-5mm}
\end{figure}

A key underlying technology that is essential for all of the above is {\em transmit  beamforming}, where signals from multiple antenna elements combine constructively at the receiver. Consider a multi-antenna radio, with each of these antenna elements having a specific directional radiation pattern, referred as a \emph{beam}. The beams from transmitter and receiver antennas are steered to initiate communication via {\em beamforming}~\cite{Vook_2014}. The communication link is then established through the periodic {\em beam sweeping} and {\em beam selection}~\cite{Giordani_2019}. Beamforming increases the signal strength at the receiver, which in turn raises the capacity limit, mitigates interference by avoiding undesirable signals at neighboring receivers, and combats the effect of pronounced path loss at high frequencies. Thus, beamforming is considered as a critical component of all modern WiFi~\cite{Nitsche2014IEEEWi-Fi} standards and is steadily being integrated into 5G~\cite{5g_nr}. 

Our survey is motivated by this observation, and we strive to answer the following two questions: (i) are there fundamental limitations of traditional RF-only beamforming technology that will impact future standards evolution, and (ii) how can new data types (beyond RF) be harnessed in the future, and, given the possible information explosion by acquiring such multimodal sensor feeds, can they be analyzed through emerging machine learning methods to guide real-time beamforming decisions? To ensure a focused discussion, we emphasize use-cases that will shape the future standards beyond 5G (henceforth referred to as \emph{NextG}), namely, beamforming scenarios that combine very large number of antenna elements and mobility. 
As an indicative example of a mmWave vehicular network that we cover in this survey, Fig.~\ref{fig_overview} shows moving vehicles beamforming towards a static base station by combining data from RF and non-RF modalities, and then using ML to identify a smaller set of beam-pairs for further optimization, instead of an exhaustive search. We begin our discussion by highlighting the need for beamforming with massive number of antennas and the use of AI/ML in beamforming communication systems.


\noindent
$\bullet$ {\bf Need for Beamforming in NextG Standards:} 
The 5G New Radio (5G-NR) standard provisions for use of both sub-6 GHz as well as millimeter wave (mmWave) frequency bands from 24.25 GHz to 52.6 GHz~\cite{5g_nr}. The sub-6 GHz band is already congested, and this problem worsens when a large data transfer needs to occur at short contact times, typically seen in mobile environments with few antennas~\cite{Giordani_2019}. While mmWave-band transmission increases capacity using wider bandwidth (up to 2 GHz)~\cite{Hemadeh_2018}, it also suffers from severe attenuation and penetration loss~\cite{Giordani_2019}. Phased-array antennas~\cite{Mailloux_2017} address the attenuation problem by leveraging the highly directional gain of the antenna elements, thereby focusing radiated RF energy into beams. This capability is enhanced in higher frequencies given the dense packing of antenna elements, i.e., higher order phased arrays are possible with proportional increase in the number of beams. While theoretically hundreds of antenna elements can be packed in a $1$cm $\times$ $1$cm area for mmWave band operation, the bottleneck lies in the complexity of processing methods and the computational resource available to properly configure the beams. Even though it is economically feasible to create large phased arrays, scaling beyond 8-12 antennas   while supporting real-time operation in small form factor wireless devices still remains an open challenge. Thus, there is need to re-visit existing approaches to beamforming to potentially scale up to thousands of antenna elements, as is being envisaged in NextG standards~\cite{Zhang_2020}.  
\pgfplotsset{grid style={dashed}}
\begin{figure}[hbtp]
	\vspace*{-0.15in}
	\centering
{\resizebox{0.35\textwidth}{!}{\begin{tikzpicture}[thick,scale=.80, every node/.style={scale=.80}]
				\begin{axis}[
				grid=major,
				xmin = 2010,
				xmax = 2020,
				xtick = {2010, 2011, 2012, 2013, 2014, 2015, 2016, 2017, 2018, 2019, 2020},
				xticklabels = {2010, 2011, 2012, 2013, 2014, 2015, 2016, 2017, 2018, 2019, 2020},
				ymin=0,
				ymax=200,
				scaled ticks=false,
				x tick label style={rotate=45,anchor=east},
				x label style={at={(axis description cs:0.5,-0.08)},anchor=north},
				axis x line=bottom,
				axis y line=left,
				axis line style=-,
				minor tick style={draw=none},
				xlabel= Year of Reference,
				ylabel= Number of Articles,
                smooth,
				cycle list={
				mark=*, blue,thick, solid\\
				mark=triangle*, magenta,thick, solid\\
				},
				every axis legend/.append style={xshift=-10pt}
				]
				\addplot plot coordinates{(2010, 3)(2011, 7)(2012, 5)(2013, 7)(2014, 19)(2015, 25)(2016, 31)(2017, 50)(2018, 75)(2019, 117)(2020, 170)};
				\addplot plot coordinates{(2010, 0)(2011, 0)(2012, 0)(2013, 3)(2014, 6)(2015, 8)(2016, 20)(2017, 36)(2018, 62)(2019, 80)(2020, 92)};
				\legend{mmWave + Beamforming, 5G + Beamforming}
				\end{axis}
				\end{tikzpicture}}}
		\vspace*{0.0in}
        \caption{Number of articles referencing beamforming in 5G for mmWave.}
	\label{fig_trend}
	\vspace{-5mm}
\end{figure}
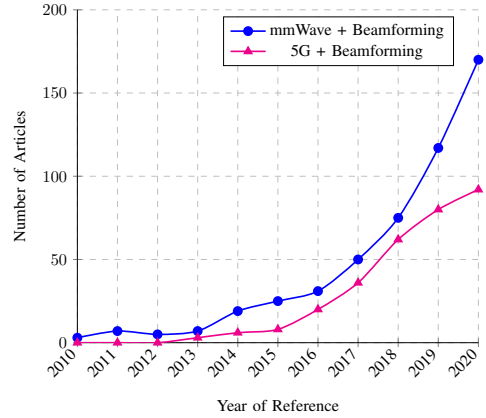

\noindent
$\bullet$ {\bf Beamforming with Massive Number of Antenna Elements:} While the possibility of having multiple antennas resulted in the multiple input and multiple output (MIMO) paradigm in WiFi networks starting with 802.11n, these were limited to 4x4 MIMO. Massive MIMO (mMIMO) scales this number up to the order of hundreds of antennas at least at the base station (BS) side, offering great flexibility in forming single directional beams~\cite{Zhang_2015} as well as multi-beams that can simultaneously target multiple users in what is known an multi-user MIMO (MU-MIMO)~\cite{Li_2020}.   
The implementation of mMIMO depends on acquiring accurate channel state information (CSI), which is then used to compensate for wireless channel distortions in a method called as {\em precoding}. The legacy CSI schemes inherited from 4G LTE consists of either: (a) codebook-based feedback for frequency division duplex (FDD) operation, or (b) reciprocity-based sounding for the time division duplex (TDD) operation at the base station~\cite{Ghosh_2010}. The higher number of antenna elements in mMIMO increases the complexity of optimal codebook design, while the overhead of CSI measurement from downlink pilot and feedback also increases at the user devices. For these reasons, this survey focuses on ways to facilitate beamforming for mMIMO, going beyond CSI-based methods.
Indeed, the many degrees of freedom in codebook design and then active selection of the actual code (i.e., selection of the beam as decided by that code) for a given situation make optimal deterministic and even most heuristic based solutions infeasible for deployment. 

\noindent
$\bullet$ {\bf Motivation for using AI-enabled Beamforming:}
Artificial intelligence and machine learning (AI/ML) based algorithms have been effectively demonstrated to outperform classical approaches in wireless-centric tasks of modulation recognition~\cite{Soltani_2019}, RF fingerprinting~\cite{Sankhe_2020}, rogue transmitter detection~\cite{Roy_2020}, etc. The use of AI-enabled algorithms to solve the above mentioned beamforming in nextG networks is still in a nascent stage. The general approach so far on using ML involves RF channel estimation followed by channel equalization by using different neural network-based architectures that accept a stream of in/quadrature phase (I/Q) samples collected by the receiver. We believe there is a vast untapped potential for AI-enabled techniques for extracting relevant information using different types of  modalities, for e.g., images can recognize the location of the target BS and this can rapidly reduce the number of candidate beams to be explored.  We refer to this emerging research trend in the domain of out-of-band beamforming as {\em multimodal beamforming}.

\noindent
$\bullet$ {\bf Scope of this Survey:} 
The statistics presented in Fig.~\ref{fig_trend}, comprise of the number of articles (including patents), from Google Scholar search results, that have referenced the terms {\em beamforming in 5G} and {\em beamforming in mmWave}. 
We believe this survey will serve the wireless research community working on beamforming in high frequency band (30-300 GHz), as in these frequencies, beamforming lies on the critical path to combat signal attenuation. We introduce and analyze the notion of multimodal beamforming for mmWave frequencies by recognizing the existing interest in the intersection of MIMO systems, wireless AI/ML and the NextG bandwidth needs. Furthermore, we emphasise the vehicular scenario shown in Fig.~\ref{fig_overview}, as it poses challenges caused by mobility that cannot be addressed in feasible time-scales through legacy methods for such large beamforming antenna arrays. As evidence of community interest on this general theme, we see a spike in citations (88 citations within 2 years) for the publicly available dataset called Raymobtime~\cite{Klautau_2018}), which contains multimodal non-RF sensor data alongwith the corresponding RF ground truths for the purpose of mmWave beamforming in a V2X environment. 

\textcolor{black}{While we strive to produce a comprehensive survey on this subject matter, we skip the reviews on} the basics of mmWave channel models, mMIMO, different beamforming system models and techniques, as there exist plethora of survey literature focusing on these fundamentals, \textcolor{black}{and is out-of-scope considering our focus area.} {\color{black} For example, the promise of mmWave communication in 5G is extensively reviewed in~\cite{Niu_2015}, the use of mmWave band for vehicular communication is surveyed in~\cite{Ghafoor_2020}, applications of mMIMO are surveyed in~\cite{Lu_2014} and~\cite{Zheng_2015}, detailed analysis of general RF-only beamforming in indoor and outdoor mmWave communications can be found in~\cite{Kutty_2016}. RF-only beamforming can have digital and analog beamforming, as well as hybrid approaches that combine the two. Related models and system architectures that contrast these three approaches are described in~\cite{Molisch_2017} and~\cite{Ahmed_2018}.} A flow-graph summarizing the existing surveys related to the ``beamforming in 5G/NextG" systems is shown in Fig.~\ref{fig:survey_organization}, and we explore each of these topics in their relevant sections later in this paper. We broadly categorize the trend of existing surveys \textcolor{black}{on that topic in three groups}: beamforming techniques for 5G, hybrid beamforming, and out-of-band beamforming; where the first two categories are related to the traditional beamforming \textcolor{black}{process}, \textcolor{black}{while} the last one is aligned towards \textcolor{black}{out-of-the-box solutions.} 
In this regard, the purpose of this survey is to identify the shortcomings in the traditional beamforming methods and \textcolor{black}{identify the advantages of using non-RF modalities}
to facilitate the beamforming process, \textcolor{black}{considering} nextG communications. \textcolor{black}{Ultimately,} we make a case for expanding the research focus towards incorporating such non-RF sensor modalities in combination with AI/ML, as a feasible pathway for NextG networks.

\usetikzlibrary{arrows,shapes,positioning,shadows,trees}

\tikzset{
  basic/.style  = {draw, text width=3cm, text = white, drop shadow, font=\sffamily, rectangle},
  root/.style   = {basic, rounded corners=2pt, thin, align=center,
                   fill={rgb:red,0;green,3;blue,5}},
  level 2/.style = {basic, rounded corners=6pt, thin,align=center, fill={rgb:red,148;green,103;blue,189},
                   text width=9em},
  level 3/.style = {basic, rounded corners=2pt, thin, align=center, fill={rgb:red,127;green,127;blue,127}, text width=8.5em},
}


\begin{figure*}[hbtp]
	\vspace*{-0.15in}
	\centering
{\resizebox{0.62\textwidth}{!}{\begin{tikzpicture}[
  level 1/.style={sibling distance=50mm},
  edge from parent/.style={->,draw}, >=latex, line width=0.6mm]

\node[root] {Beamforming in 5G/NextG Surveys}
  child {node[level 2] (c1) {5G Networks}}
  child {node[level 2] (c2) {Hybrid Beamforming}}
  child {node[level 2] (c3) {Out-of-band Beamforming}};

\begin{scope}[every node/.style={level 3}]
\node [below of = c1, xshift=35pt, yshift=-5pt] (c11) {mmWave Channel Models~\cite{Hemadeh_2018}};
\node [below of = c11, yshift=-5pt] (c12) {mMIMO~\cite{Lu_2014, Zheng_2015}};
\node [below of = c12, yshift=-10pt] (c13) {mmWave Communication~\cite{Niu_2015, Ghafoor_2020, Kutty_2016}};

\node [below of = c2, xshift=20pt, yshift=-15pt] (c21) {Architecture and System Models~\cite{Ahmed_2018}};
\node [below of = c21, yshift=-10pt] (c22) {mMIMO~\cite{Molisch_2017}};
\node [below of = c22, yshift=-5pt] (c23) {mmWave beamforming~\cite{Roh_2014}};
\node [below of = c23, yshift=-5pt] (c24) {Wireless Energy Harvesting~\cite{Alsaba_2018}};

\node [below of = c3, xshift=20pt, yshift=-10pt, fill={green!50!black}] (c31) {Out-of-band RF-based (Sec.~\ref{sec:out-of-band-RF-based})};
\node [below of = c31, yshift=-10pt, fill={green!50!black}] (c32) {Non-RF based (Sec. \ref{subsec:dataAcq_and_pre} - \ref{subsec:single_non_RF})};
\node [below of = c32, yshift=-12pt, fill={green!50!black}] (c33) {Multimodal (Sec.~\ref{subsec:multimodal_beamforming})};
\node [below of = c33, yshift=-17pt, fill={green!50!black}] (c34) {AI-enabled Implementations (Sec.~\ref{sec:beamformingNonRF} - \ref{sec:proposed_work})};
\end{scope}

\foreach \value in {1,2,3}
  \draw[->] (c1.195) |- (c1\value.west);

\foreach \value in {1,...,4}
  \draw[->] (c2.195) |- (c2\value.west);

\foreach \value in {1,...,4}
  \draw[->] (c3.195) |- (c3\value.west);
\end{tikzpicture}}}
        \caption{Existing surveys on different areas of beamforming for 5G and beyond. This survey mainly focuses on use of {\em out-of-band} beamforming in recent literature for the 5G and NextG networks. We describe different categories of out-of-band beamforming in the corresponding sections in the rest of the paper.}
	\label{fig:survey_organization}
\end{figure*}
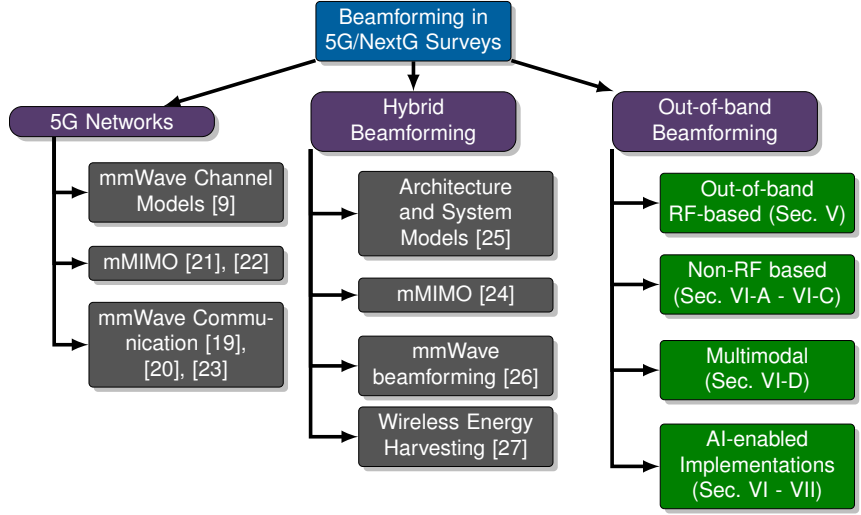


\noindent
$\bullet$ {\bf Organization of this Survey:} 
The remainder of this article is organized as follows. High level differences between different traditional and non-RF based beamforming techniques for nextG networks are described in  Sec.~\ref{sec:rf_and_non_rf_beamforming}, followed by a comprehensive review of published surveys in related areas of beamforming in Sec.~\ref{sec:survey_of_surveys}. 
The drawbacks in the legacy signal processing approaches in the RF domain lead us towards the use of new tools, such as ML-based approaches, which we describe in  Sec.~\ref{sec:aid_to_beamforming}. We then turn towards the use of out-of-band RF and and non-RF data for beamforming is presented in Sec.~\ref{sec:out-of-band-RF-based}, and Sec.~\ref{sec:beamformingNonRF}. With multiple data modalities available, we present few novel fusion techniques for fast beam selection in Sec.~\ref{sec:proposed_work}. We discuss different application areas of multimodal beamforming in Sec.~\ref{sec:applications} with additional emerging trends included in Sec.~\ref{sec:emerging_trend}. The conclusions are drawn in the last section. Acronyms used in the survey are listed in Tab.~\ref{tab:acronyms}.


\begin{table}
    \centering
    \resizebox{0.4\textwidth}{!}{\begin{tabular}{c|l} 
    \specialrule{.2em}{.1em}{.1em} 
    \rowcolor{lightgray}
    {\bf Acronyms} & {\bf Meanings} \\ 
    \specialrule{.2em}{.1em}{.1em} 
    2D & Two Dimensional \\
    3D & Three Dimensional \\
    5G & 5th Generation \\
     5G NR & 5th Generation New Radio \\
     6G & 6th Generation \\
     AI & Artificial Intelligence \\
     AoA & Angle of Arrival \\
     AP & Access Point \\
     AR & Augmented Reality \\
     BBS & Blind Beam Steering \\
     BS & Base Station \\
     CDF & Cumulative Distribution Function \\
     CNN & Convolutional Neural Network\\
     CRN & Cognitive Radio Network \\
     CS & Compressive Sensing \\
     CSI & Channel State Information \\
     DL & Deep Learning \\
     DoA & Direction of Arrival \\
     FDD & Frequency Division Duplex \\ 
    EH & Energy Harvesting \\
    EHF & Extremely High Frequency \\
    FANET &  Flying Ad-hoc NETworks \\
    FML & Fast Machine Learning \\
    IA & Initial Access \\
    ID & Identification \\
    IoT & Internet of Things \\
    IR & Infrared \\
    LiDAR & Light Detection and Ranging \\
    LMMSE & Linear Minimum Mean Squared Estimation \\
    LSTM & Long Short-Term Memory \\
    MAB & Multi Armed Bandit\\
    MCS & Modulation and Coding Scheme\\
    MIMO & Multiple Input Multiple Output\\
    ML & Machine Learning \\
    MLP & Multi-layer Perceptron \\
    MMSE & Minimum Mean Squared Estimation \\
    mMIMO & massive MIMO \\
    MU-MIMO & Multiple User MIMO \\
    mmWave & millimeter Wave \\
    MR & Mixed Reality \\
    MSE & Mean Squared Error \\
    nextG & Next Generation \\ 
    NOMA & Non-Orthogonal Multiple Access\\
    NLoS & Non Line of Sight \\
    OFDM & Orthogonal Frequency Division Multiplexing\\
    RADAR & Radio Detection And Ranging \\
    RF & Radio Frequency \\
    RGB & Red Green Blue \\
    RMSE & Root Mean Squared Error  \\
    RS & Reference Signal \\
    RSU & Road Side Unit \\
    RX & Receiver \\
    LoS & Line of Sight \\
    SE & Spectral Efficiency \\
    SHF & Super High Frequency \\
    SINR & Signal-to-interference Noise Ratio \\
    SISO & Single Input Single Output \\
    SLAM & Simultaneous Localization And Mapping \\
    SS & Synchronization Signals \\
    TDD & Time Division Duplex \\
    THz & Terahertz \\
    TX & Transmitter \\
    UAV & Unmanned Aerial Vehicles \\
    UE & User Equipment \\
    UGV & Unmanned Ground Vehicles \\
    V2I &  Vehicle to Infrastructure \\
    V2X & Vehicle to Everything \\
    VR & Virtual Reality \\
    WID & Wireless Infrastructure Drone \\
    WiGig & Wireless Gigabit \\
    XR & eXtended Reality \\
    \specialrule{.2em}{.1em}{.1em} 
    \end{tabular}}
    \caption{The details of the frequent acronyms used in the article.}
    \label{tab:acronyms}
\end{table}

\section {Beamforming Techniques}
\label{sec:rf_and_non_rf_beamforming}


\textcolor{black}
In this section, we first analyze \textcolor{black} {the state-of-the-art in traditional beamforming techniques that may impair inclusion in future NextG standards operating in mmWave bands. We then explore the current research on non-RF based beamforming to  motivate our intent of using these methodologies to address the shortcomings of traditional RF-only beamforming.}


\subsection{Traditional Beamforming}
\textcolor{black}{Existing RF-based beamforming approaches (analog, digital, hybrid) have their unique advantages, and are applicable in specific scenarios. Indeed, the 5G-NR standard supports all three types of beamforming in the time domain~\cite{Kutty_2016}.}

A brief comparison study for these approaches is presented in Tab.~\ref{tab:beamforming_comparision}. Digital beamforming improves the spectral efficiency (SE) of a MIMO system by simultaneously transmitting data to multiple users. However, it needs a distinct RF chain per antenna, making it less cost-effective for  higher order of antenna elements. \textcolor{black}{This is one core reasons why there are  few off-the-shelf mmWave radios~\cite{piradio} which support digital beamforming even with low order (1$\times$4) of antenna elements. Unlike its digital counterpart, analog beamforming creates the beam using one element per set of antenna.} Once the best beam, among all possible combinations of beam-pairs is identified, it is activated to mitigate the impact of high pathloss in mmWave band. This is why most of the off-the-shelf mmWave devices~\cite{ni_radios, terragraph, imb_paam_conference} support only analog beamforming. Also, analog beamforming is considered mandatory in 5G-NR~\cite{Li_2020} for  mmWave operation. 

\textcolor{black}{Hybrid beamforming, on the other hand, is a combination of analog and digital beamforming. The idea of hybrid beamforming revolves around trading-off the hardware cost for the overhead of time involved in beam selection. Here, a subset of antennas is connected to a particular RF chain, as opposed to having individual RF chains for each antenna element in digital beamforming.} Even though, hybrid beamforming promises faster communication with higher order antenna elements, this is still an area on ongoing research~\cite{ibm_paam}. Additionally, for hybrid beamforming, the continuous {\em beam management} technique in a mobile environment 
involves periodic overhead~\cite{Li_2020}. Here, beam selection is done after the measurement of reference signals (RS) received in a specific direction by manipulating the beamforming weights applied across different antenna elements. 

\begin{table}[hbtp]
    \centering
\resizebox{0.48\textwidth}{!}{
     \begin{tabular}{c||c|c|c} 
     \specialrule{.2em}{.1em}{.1em} 
     \rowcolor{lightgray}
    {\bf Metrics}     & {\bf Analog} & {\bf Digital} & {\bf Hybrid} \\ \specialrule{.2em}{.1em}{.1em} 
    Degree of Freedom & Limited           & Highest         & High\\     \hline
    Implementation    & Phase Shifter     & ADC/DAC, mixers & Everything \\     \hline
    Architecture      & Simple            & Complex         & Complex \\ \hline
    Baseband Chains   & Less              & Highest         & High \\ \hline
    Complexity        & Less complex      & Complex         & Complex\\     \hline
    Power Consumption & Less              & Highest         & High \\     \hline
    Cost              & Less              & Highest         & High \\     \hline
    Inter-user Interference & High        & Lowest          & Low\\     \hline
    MIMO Support      & No                & Yes & Yes\\ \hline
    Flexibility & Fixed delay & Flexible weight vector& Flexible weight vector\\
    \specialrule{.2em}{.1em}{.1em} 
    \end{tabular}}
    \caption{Comparison of different beamforming types.}
    \label{tab:beamforming_comparision}
    
\end{table}

\begin{figure}[hbtp]
    \centering
    \includegraphics[width=0.67\linewidth]{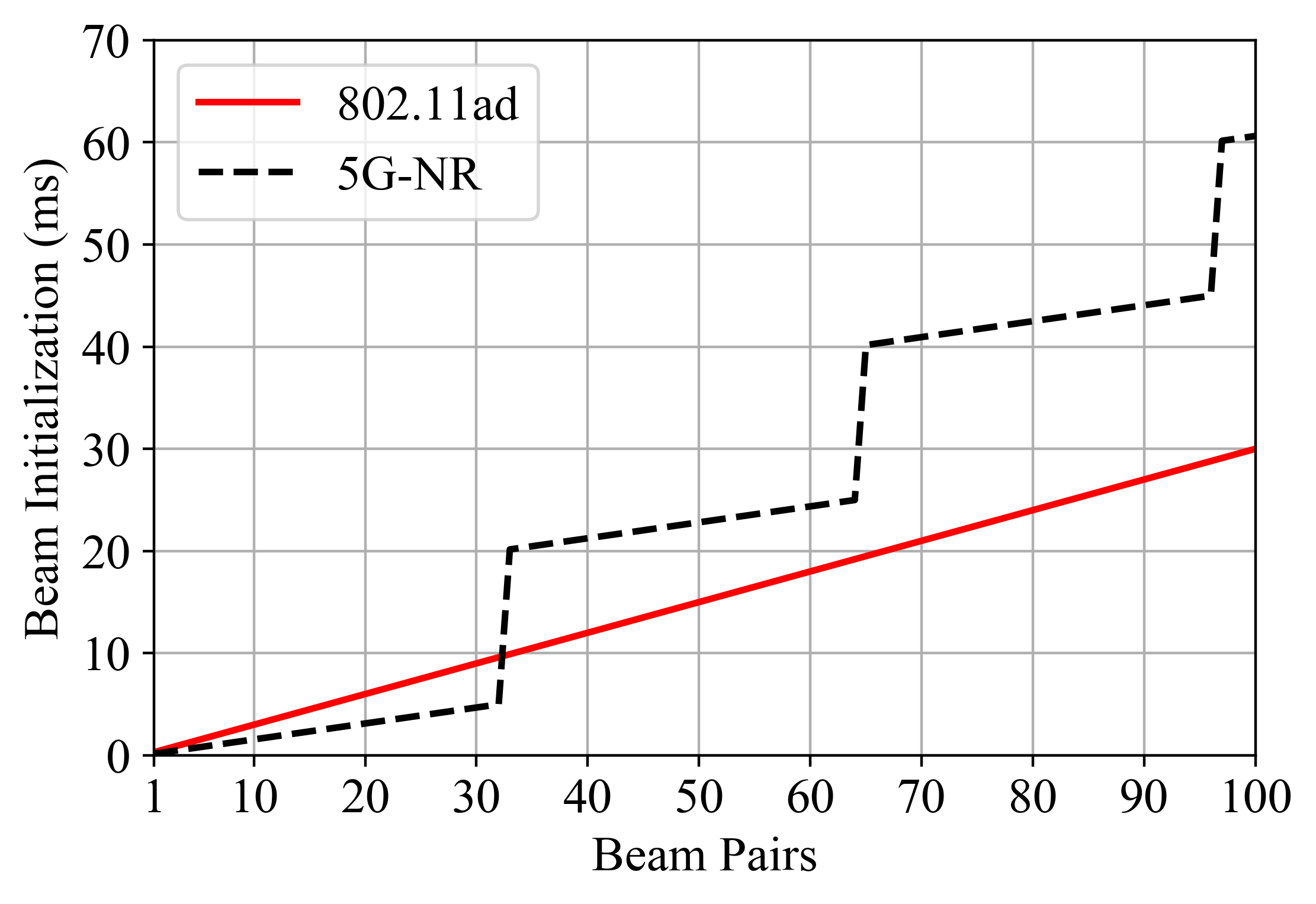}
    \caption{The average beam initialization overhead of IEEE 802.11ad and 5G-NR standards with respect to different beam search spaces. The beamforming time significantly increases with increasing number of beam pairs.}
    \label{fig:time_cost}
    \vspace{-5mm}
\end{figure}

\begin{figure*}[hbtp]
    \centering
    \includegraphics[width=0.7\linewidth]{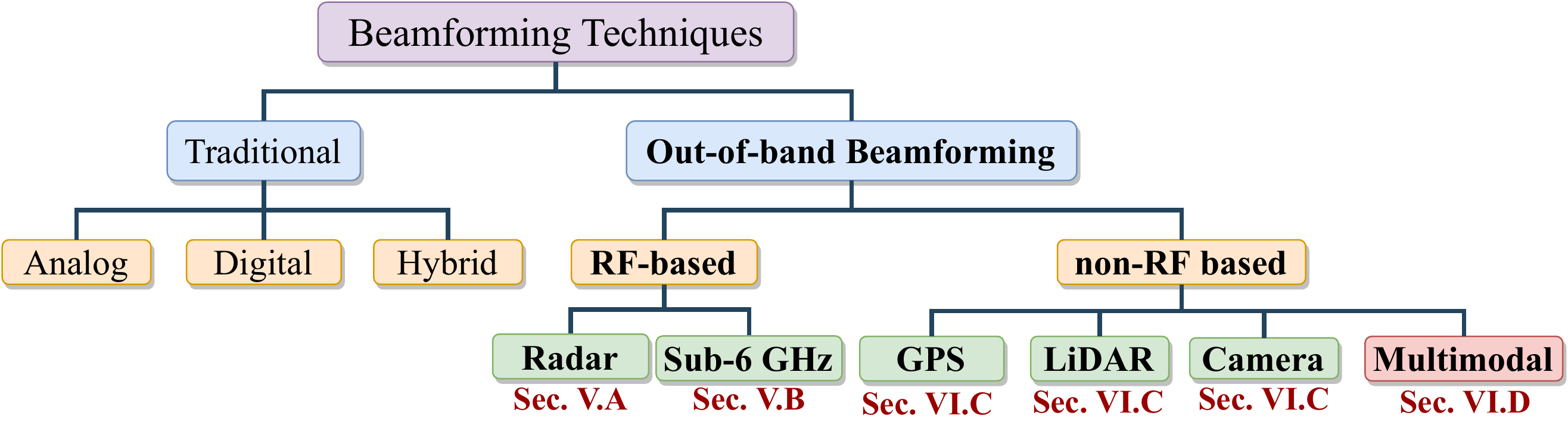}
    \caption{ The general hierarchy of different beamforming techniques, with a focus on out-of-band ones in both RF and non-RF domain. 
    The `multimodal' beamforming is highlighted and we cover novel fusion techniques for multimodal beamforming later in Sec.~\ref{sec:proposed_work} as well.} 
    \label{fig:out_of_band_bf}
    \vspace{-5mm}
\end{figure*}

\subsection{Out-of-band Beamforming}
\textcolor{black}{As discussed before, analog beamforming technique involves time-overhead of beam selection due to exhaustive search among all possible transmitter-receiver (TX-RX) antenna elements. The decision is made based on a combination of RF measurements, such as CSI, SNR etc., in the desired frequency band of communication. 
This overhead (Fig.~\ref{fig:time_cost}) is exacerbated in the case of mobile users where the position of user equipments (UEs) are changing continuously, resulting in the exhaustive search being instantiated multiple times within a few seconds. Furthermore, the wireless channel varies 10x faster at 30~GHz as opposed to 3~GHz, even for the same UE mobility rate. This results in 10x more frequent beam sweeping and channel estimation~\cite{Bjornson_2019}. 
Thus, we believe  that {\em out-of-band} RF measurements and the use of environmental non-RF data appear offer an  attractive alternative towards minimizing the overhead of exhaustive search. We refer to such approaches as {\em out-of-band beamforming} techniques. A visual representation of the existing traditional and out-of-band beamforming techniques are given in Fig.~\ref{fig:out_of_band_bf}}.




\section{Beamforming Surveys on mMIMO for 5G and Beyond}
\label{sec:survey_of_surveys}
We now review the state-of-the-art in research and analysis on mMIMO, beamforming, and mmWave communication in 5G and nextG wireless networks. The fundamentals of mMIMO and mmWave operation and the applications of mMIMO are comprehensively surveyed in earlier works~\cite{Lu_2014}\cite{Zheng_2015}. The promise of mmWave communication in 5G is extensively reviewed in~\cite{Niu_2015}, and the use of mmWave band for vehicular communication is surveyed in~\cite{Ghafoor_2020}.

From Fig.~\ref{fig_trend} we see that the research interest on beamforming in mmWave band and 5G standards are strongly coupled, as the advancements in the former are essential to meet operational requirements for the latter.
Additionally, exploration of new spectrum, assigning more bandwidth, carrier aggregation, inter-cell interference mitigation techniques, integration of mMIMO antennas, etc., are all key features that have been  extensively covered  in~\cite{Mitra_2015}. Also, the authors state that providing accessibility, flexibility, and cloud based services through proper modulation and coding scheme (MCS), mmWave and device to device (D2D) communication is the key to realize functional nextG networks. Authors in~\cite{Roh_2014} validate the notion that beamforming has a bigger role to play in mmWave bands, as compared to low frequency bands. Hence, there is great interest in beamforming optimization in mmWave bands for  nextG standards.

For sake of completeness, we mention below surveys that describe beamforming advancements tailored for sub-1 GHz, sub-6 GHz as well as sub-30 GHz 5G bands. Authors in  \cite{Rumyancev_2019} focus on the frequency allocation, beamforming techniques and custom-designed integrated circuits for those specific bands. Kutty {\em et al.}  \cite{Kutty_2016}  capture the evolution of different beamforming techniques in the context of mmWave communication. They describe different radio frequency system design and implementation for millimeter wave beamforming for indoor and outdoor communication scenarios. The authors describe the mmWave propagation characteristics in-terms of path loss and clustered multi-path structures, dominant LoS component, wideband communication and 3D spatio-temporal modeling. They also illustrate different phased array antenna architectures to support MIMO capability in mmWave beamforming. 
\textcolor{black}{Finally, the authors concur that using hybrid beamforming in the mmWave band for MIMO to minimizing cost and power consumption has great promise.}

\textcolor{black}{In a survey on hybrid beamforming for mMIMO, Molisch {\em et al.}~\cite{Molisch_2017} analyze the trade-offs of using instantaneous or average (second-order) CSI in hybrid beamforming. Here, the authors evaluate current research on various types of hybrid multiple-antenna transceivers and consider how the channel sparsity in the mmWave band can be leveraged for optimizing channel estimation and beam training. However, to get broader aspects of hybrid beamforming, we review an extensive survey by Ahmed {\em et al.} in~\cite{Ahmed_2018}, which thoroughly track the progress in this domain till 2017. In this paper the authors present different architectures of hybrid beamforming and the techniques for optimization of phase shifters, DAC/ADC resolutions and antenna configurations. From the system model perspective, they examine eight variations of hybrid beamforming and identify many resource management aspects, particularly in beam management, MAC protocol variants, which can impact the performance of hybrid beamforming. }

\textcolor{black}{Researchers have also done surveys on beamforming in cognitive radio networks~(CRN)~\cite{Xu_2015} and wireless energy harvesting~(EH)~\cite{Alsaba_2018}. The basic principle of achieving high SINR through beamforming makes it a potential candidate for transmitters within EH networks. The efficacy of using CRN and EH for energy-constrained communication networks has been exhibited in~\cite{Hossain_2014, Raghunathan_2006, Pentikousis_2010, Chen_2011}, along with focused research on the topics of military communications and submarines~\cite{Alsaba_2018}, sensor networks~\cite{Sharma_2010},~\cite{mohanti2018wifed}, and medical implants~\cite{Zhang_2009}, \cite{Banou2019BeamformingCommunication}. In~\cite{Alsaba_2018}, the authors also advocate for beamforming for CRN in nextG networks. Tab.~\ref{tab:survey_comparision} summarizes the published surveys related to beamforming in the mmWave frequency band. 
}
However, the goal of this survey paper is to describe  beamforming techniques that  exploit out-of-band RF and non-RF multimodal data for nextG networks. To motivate the case for out-of band RF and multimodal data, we first identify the limitations of traditional beamforming methods using RF-only data.






\begin{table*}[hbtp]
    \setlength{\tabcolsep}{5pt}
    \centering
\resizebox{1.0\textwidth}{!}{
     \begin{tabular}{c|c|p{1.8cm}|p{2cm}|p{1.5cm}|p{2cm}|p{2cm}|p{2cm}} 
     \specialrule{.2em}{.1em}{.1em} 
     \rowcolor{lightgray}
    Paper & Year &  Beamforming in mmWave & Inclusion of 5G NR Standard & Supports MIMO & Shortcomings of Traditional mmWave Beamforming & Aiding the Traditional mmWave Beamforming & non-RF Data for Beamforming\\\specialrule{.2em}{.1em}{.1em} 
     \cite{Rumyancev_2019}& 2019 & \cmark & \xmark &  \xmark & \xmark &\xmark & \xmark\\ \hline 
     \cite{Kutty_2016}& 2016 & \cmark & \xmark & \cmark & \xmark& \xmark&\xmark \\ \hline 
     \cite{Molisch_2017}& 2017 & \cmark & \xmark& \cmark &\xmark & \xmark&\xmark \\ \hline 
     \cite{Ahmed_2018}& 2018 & \cmark & \cmark & \cmark & \cmark &\xmark & \xmark\\ \hline 
     This & 2021 & \cmark & \cmark & \cmark & \cmark & \cmark & \cmark \\\specialrule{.2em}{.1em}{.1em} 
    \end{tabular}}
        \caption{Comparative analysis of existing surveys for beamforming in mmWave band.}
            \label{tab:survey_comparision}
\end{table*}



\section{Limitations of Traditional RF-only based Approaches}
\label{sec:aid_to_beamforming}
The traditional RF-only beamforming approach utilizes one of these two options: (a) estimate the mmWave channel at the receiver, and send this information back to the transmitter for generating the precoding weights, (b) sweep through the antenna codebook elements of the transmitter and receiver. In this section, we discuss how these RF-only solutions for beamforming impose significant overhead for mmWave links. For the first option, we discuss the published literature related to continuous channel estimation and closed loop feedback to the transmitter for beamforming in mmWave mMIMO. For the second option, we review works on continuous beam sweeping and beam alignment needed for sustainable communication, and introduce the out-of-band beamforming as a solution to overcome the expensive channel estimation and beam sweeping tasks.

\subsection{Channel Estimation}
\textcolor{black}{Modern MIMO wireless communication systems use spatial multiplexing to improve the data throughput in a rich scattering environment. In order to send multiple data streams through the channel, a set of precoding and combining weights are derived from the channel matrix to recover each data stream independently. These weights contain both the magnitude and phase of the channel and are normally applied in the digital domain. Depending on the application, uplink and downlink communication is performed either in Time Division Duplex (TDD) or Frequency Division Duplex (FDD). While FDD allows for full duplex wireless interfaces, TDD is more practical in dense cellular deployments.}

\textcolor{black}{Accurate channel estimation is challenging in mmWave mMIMO systems due to the sheer magnitude of number of antennas, low SNR channels, hardware constraints, etc. The basic idea behind {\em channel estimation} is to acquire the most current CSI at the receiver and forward it to the transmitter~\cite{Belgiovine2021} within some application specific latency bound so as to allow both the {\em channel sounding} and {\em data transfer} phases to be completed within the channel coherence time. Such stringent thresholds on CSI feedback latency ensure that the transmitter can turn around its radio front-end and leverage channel reciprocity for the {\em downlink transmission}. CSI is also key for realizing spatial multiplexing, where independent paths are available in the channel between the transmitter and receiver. Typically, perfect channel estimation is assumed in literature, which is hard to achieve in practical real-life deployments~\cite{Belgiovine2021}. While channel estimation is typically accomplished via classical methods, in recent years, ML based methods are being researched to overcome the limitations of these classical approaches. We next summarize the fundamental differences between these two approaches.}

\subsubsection{Traditional Channel Estimation}
\textcolor{black}{Least square (LS) estimation is one of the simplest and  fastest channel estimators, although its performance is affected by high mean squared error (MSE) at low SNR levels. In such cases, an additional filtering based on minimum mean squared estimation (MMSE), typically a linear MMSE (LMMSE)~\cite{Savaux2017}, is adopted to improve the LS estimation by filtering out the noise. Even in its linear form, MMSE turns out to be an expensive and scales poorly~\cite{Savaux2017}. This calls for further research on more efficient methods for mMIMO systems. }
\subsubsection{ML-based Channel Estimation}
ML and, in particular, deep learning (DL) are increasingly considered for channel estimation in many areas of wireless communication~\cite{Belgiovine2021}.
\begin{table*}[t]
\centering
\begin{tabular}{c|c|p{0.5cm}|p{6cm}|p{1cm}|p{2.5cm}} 
     \specialrule{.2em}{.1em}{.1em} 
\rowcolor{lightgray}
Method                  & Type of DL Model & $L$  & Inference  Complexity (Forward Step)                                             & OFDM & Additional Comments \\
\specialrule{.2em}{.1em}{.1em} 
Huang {\em et al.}~{\cite{Huang2018}} & MLP              & 6  & $\mathcal{O}(\sum_{l=1}^{L}N_lI_l + \mathcal{G})$            & No    &        $K$ models needed to operate on OFDM             \\ \hline
Dong {\em et al.}~{\cite{Dong2019}}       & CNN              & 10 & $\mathcal{O}(K\mathcal{T} + N_TN_R\sum_{l=1}^{L}F_lN_{l-1}N_l)$ & Yes\textsuperscript{\textdagger}    &                     \\\hline
He {\em et al.}~{\cite{He2018}} &
  LDAMP + CNN &
  10 &
  $\mathcal{O}(\sum_{l=1}^{L} \mathcal{L} + L\sum_{c=1}^{20} W_cH_cF_c^2N_{c-1}N_c)$ &
  No & $K$ models needed to operate on OFDM 
  \\\hline
Balevi {\em et al.}~{\cite{Balevi2020}} &
  CNN + upsampling &
  6 &
  $\mathcal{O}(E (W_1H_1N_0N_k + \sum_{l=2}^{L}2W_{l-1}2H_{l-1}N_{l-1}N_k))$ &
  Yes\textsuperscript{\textdagger} &
  $E$ has no upper bound \\\hline
Belgiovine {\em et al.}~\cite{Belgiovine2021}                & MLP              & 3  & $\mathcal{O}(\sum_{l=1}^{L}N_lI_l)$                             & Yes\textsuperscript{\textsection}    &     \\                 
     \specialrule{.2em}{.1em}{.1em} 
\end{tabular}
\caption{ A coarse computational complexity comparison between existing methods and proposed channel estimator. Notation: $N_T$ = number of transmitter antennas, $N_R$ = number of receiver antennas, $K$ = number of sub-carriers, $L$ = number of hidden layers, $I_i$ = number of input features of layer $i$-th, $N_i$ = number of neurons (or kernels, in case of CNNs) in $i$-th layer, $F_i$ = kernel size of $i$-th convolutional layer (assuming square kernels), $W_i$ = width of input volume for $i$-th convolutional layer, $H_i$ = height of input volume of $i$-th convolutional layer, $E$ = number of epochs, $\mathcal{L}$ = complexity of LDAMP layer (linear system) in \cite{He2018}, $\mathcal{T}$ = complexity of tentative estimation (linear system, including matrix multiplications and inversions) in \cite{Dong2019}, $\mathcal{G}$ = complexity of additional linear system needed to compute complex channel coefficients from DoA estimation (requires matrix inversion) in \cite{Huang2018}, \textdagger~= method requires OFDM demodulation, \textsection~= method does \textit{not} require OFDM demodulation.}
\label{tab:complexity-comp}
\vspace{-5mm}
\end{table*}
\textcolor{black}{An end-to-end orthogonal frequency-division multiplexing (OFDM) symbol decoding method using MLP is presented by~\cite{Ye2018} through the process of treating a single input single output (SISO) channel model as a black box. }

Applying DL based approaches for CSI estimation in mMIMO is still at a developing stage. Given the high dimensionality in mMIMO, especially when involving OFDM techniques, the majority of existing solutions use complex and deep architectures to estimate large channel matrices. These solutions treat the multi-dimensional input signal as a single entity and often require additional prior or post-estimation steps. Although use of very deep architectures is a growing trend, their  complexity usually limits use in edge devices that are typically  constrained in power and processing capability.Dong {\em et al.}~\cite{Dong2019}  use convolutional neural networks (CNN) to improve the quality of a coarse initial estimate of the channel matrix in a method 
called {\em tentative estimation}. To exploit adjacent sub-carrier frequency correlations, the coarse channel estimate matrices are concatenated in large input tensors and processed by a neural network consisting of 10 convolutional layers. He {\em et al.}~\cite{He2018} propose a 10-layer learned denoising-based approximate message passing~(LDAMP) architecture, based on the unfolding of the iterative D-AMP algorithm. As the estimated channel is treated as a noisy 2D image, each layer relies on an additional denoising CNN,  which is 20-layers deep and is used to update the channel estimated in the previous layer. \textcolor{black}{Although CNNs are  efficient in terms of number of parameters, the resulting complexity poses a challenge at the edge for the deep architectures, like those proposed in~\cite{Dong2019} and ~\cite{He2018}}. In the context of single-carrier systems~\cite{Huang2018} devises an uplink transmission for single antenna users and multi-antenna BS using a 6-layer MLP to first estimate the direction of arrival (DoA) and then determine the channel for each user, by expressing the channel estimate as a function of the DoA and solving an additional linear system of equations. Balevi {\em et al.}~\cite{Balevi2020},  describe an online training method based on deep image prior scheme, using a 6-layer architecture based on $1\times1$ convolutions and upsampling, which performs denoising of the received signal before a traditional LS estimation. Although the number of parameters here is low, this method requires the network to be trained during every transmission for thousands of epochs, without any guarantee that this \textcolor{black}{training step will be completed within the channel coherence time.} For single-carrier solutions, $K$ separate models, where $K$ is the number of sub-carriers, should be trained for deployment in practical OFDM systems, resulting additional complexity over the LS estimation.


Recently, Belgiovine {\em et al.} \cite{Belgiovine2021},  demonstrated an edge-oriented MLP with compact architecture that exploits similarities in each transmit-receive antenna pairs to estimate their channels at each sub-carrier independently.Due to the the inherently parallel nature,  DL models can complete channel estimation process with improved quality of estimation and reduced computational time.  Tab.~\ref{tab:complexity-comp} summarizes the time complexity of existing methods and compares how this DL approach results in a much simpler model that is suitable for edge architectures.

\subsection{Feedback from Receiver}
The next step after the CSI estimation is to send the feedback to transmitter from the receiver.  \textcolor{black}{If TDD is employed, then there are two phases involving the BS and UE:} \textit{channel sounding}, in which case the UE performs CSI estimation for the complete MIMO channel and sends it back to the BS, and \textit{data transfer}, where the BS uses the received CSI estimation to compute the precoder and combiner's weights for directional beams. On the other hand, FDD schemes allow for the upload of CSI to BS on a dedicated band. Transferring the entire CSI is impractical considering the available channel bandwidth, and therefore, its relative information increases linearly with the number of transmitter antennas in mMIMO systems. 
Methods based on compressive sensing (CS)~\cite{Kuo2012,Rao2014, daubechies2004iterative, donoho2009message, li2009user, metzler2016denoising} focus on reducing feedback overhead by using spatial and temporal correlation of CSI. In particular, correlated CSI can be transformed into an uncorrelated sparse vector in some cases and CS can be used to obtain a sufficiently accurate estimate of such sparse vectors. 
\textcolor{black}{However, channels are not usually sparse and may not always have an interpret-able structure.}
In order to overcome this limitation and learn a better CSI compression function, DL is used in~\cite{Wen2018} to learn an encoder-decoder scheme to compress the CSI into a lower dimensional space and transmit a compressed information of it to the BS.
Yet, CSI feedback ushers in  complexity and overhead, which must otherwise be kept as low as possible.  \textcolor{black}{This motivates the need for further studies on advanced channel feature extraction mechanisms suitable for edge devices.}



\subsection{Beam Sweeping}
The alternative of using CSI feedback for beamforming is to perform an exhaustive beam search based on transmitter and receiver codebook to establish the directional link in 5G and nextG networks.
In general, the overall link establishment process comprises of 4 different steps \cite{giordani2018tutorial}: (a) {\em beam sweeping}, which involves exploration of all the available beams through transmission/reception of reference signals; (b) {\em beam measurement}, which evaluates the quality of each beam through a predefined metric i.e. SNR; (c) {\em beam selection}, which is the process of selecting the best beam based on the beam measurement results; (d) {\em beam reporting}, which shares beam quality/decision information, usually from the UE to the BS. In this section, we briefly summarize the beam sweeping process defined by the 5G-NR standard during the initial access (IA), which assumes that no link has been previously established between the user equipment (UEs) and the base station (gNB). Notice that methods alternative to beam sweeping have attracted most of the research efforts due to the inefficiency and high overhead of legacy brute force algorithms.
The 5G-NR standard~\cite{3gpp_5gnr} defines an exhaustive beam search process to find the best beam-pair configuration between the UE and the gNB.
\textcolor{black}{For larger antenna arrays, the time required to sweep through different sectors is not scalable for time sensitive applications.}

\noindent
$\bullet$ {\bf Exhaustive Beam Search Time in 5G NR: }Consider a gNB-UE pair, with codebook sizes $C_{UE}$ and $C_{gNB}$ respectively. Then, the total number of beam directions to be scanned is $|\mathcal{C}| = C_{UE}\times C_{gNB}$. During the initial access~(IA), the gNB and the UE exchange a number of messages to find the best beam pair. \textcolor{black}{During this process,} the gNB sequentially transmits synchronization signals~(SS) in each codebook element. Meanwhile, the UE also switches among sectors to receive in different codebook elements until all $|\mathcal{C}|$ possible beam configurations are swept.
The SS transmitted in a certain beam configuration are referred to as SS blocks, with multiple SS blocks from different beam configurations grouped into one SS burst. \textcolor{black}{The SS burst duration ($T_{ssb}$)  is fixed  at $5\,ms$ in the NR standard, and}
it is typically transmitted with a periodicity ($T_p$) of $20\,ms$~\cite{barati2020energy}, although different values are supported $T_p \in \{5, 10, 20, 40, 80, 160 \}$ $ms$. In the mmWave band, a maximum of 32 SS blocks fit within a SS burst, and the number of explored beams per block is dependent on the beamforming technique. For example, hybrid and digital beamforming architectures allow transmitting or receiving multiple beams simultaneously. On the other hand, analog beamforming architectures only enable one beam per configuration, 
\textcolor{black}{requiring as many SS blocks as beam configurations to perform the beam sweeping process}. The total beam sweeping time ($T_{bs}$) can thus be expressed as:

\begin{equation}
    T_{bs} = T_p \times \left \lfloor\frac{ |\mathcal{C}| - 1}{32} \right \rfloor + T_{ssb}.
\end{equation}
\textcolor{black}{As showcased in Fig.~\ref{fig:time_cost}, the beam sweeping time increases linearly with increasing beam pairs, and this increment is larger in case of 5G-NR ($\approx40$ms for $|\mathcal{C}|>64$). Such delay has the potential for severely degrading the 5G NR performance, where time sensitive applications typically require latencies of~$\leq 10~ms$}~\cite{3gpp_5gnr}.
 

\textcolor{black}{In order to reduce the overhead of the complicated channel estimation and time-consuming beam-sweeping techniques, multiple out-of-band approaches have been explored in the recent literature, with the aim of achieving low overhead. These beamforming techniques can be broadly categorized into (a) \textit{RF-based} and (b) \textit{non-RF based}, with their different sub-categories illustrated in Fig.~\ref{fig:out_of_band_bf}.
In the next sections, we explore in detail each of these categories.}



\section{Out-of-band RF based Beamforming}
\label{sec:out-of-band-RF-based}
\begin{figure}
    \centering
    \includegraphics[trim=0cm 0cm 0cm 0cm, clip, width=0.45\textwidth]{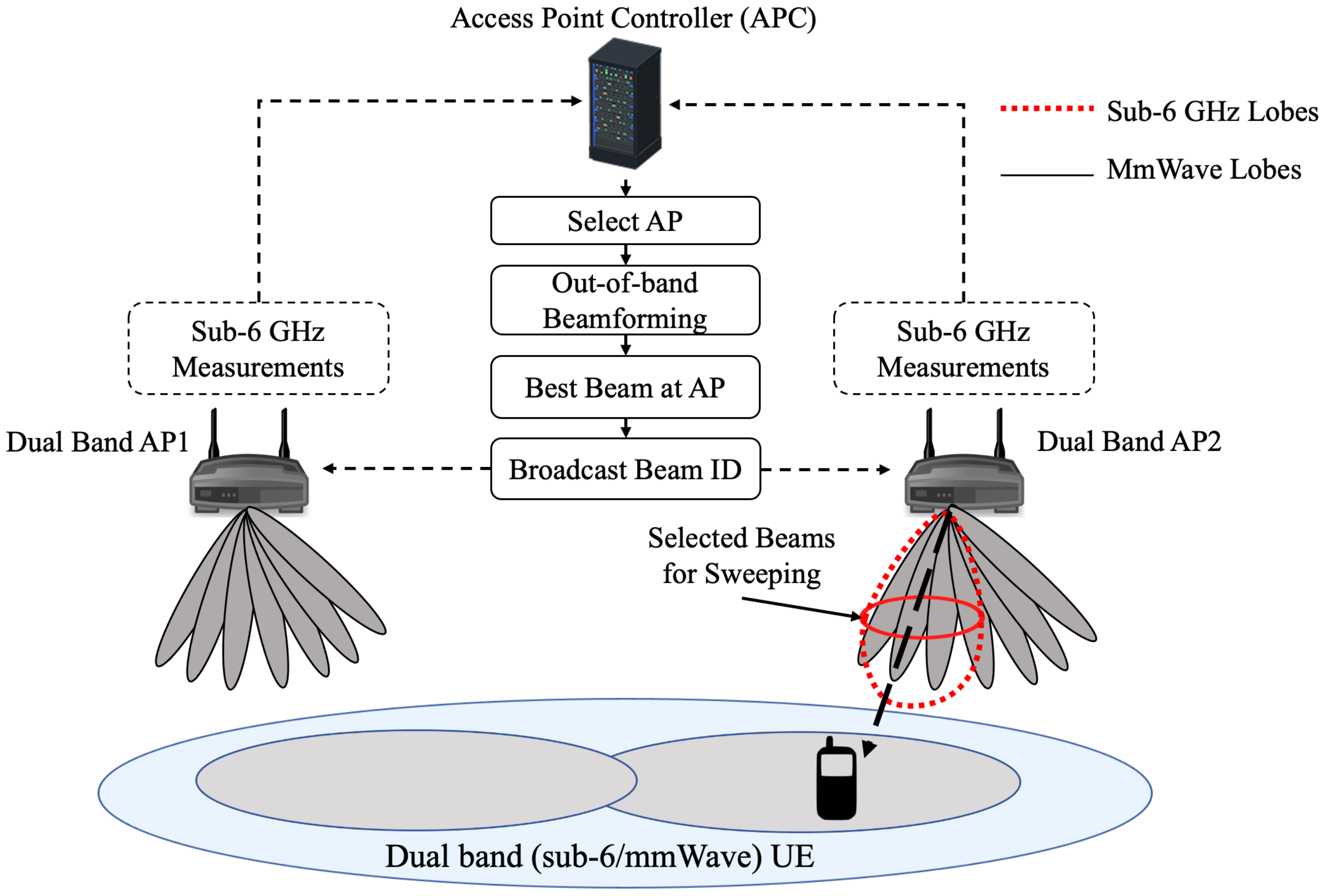}
    \caption{The AP exploits the measurements at lower frequencies, sub-6~GHz in this example, to propose a set of most likely beam pairs.}
    \label{fig:out-of-band}
    \vspace{-5mm}
\end{figure}

The main idea behind leveraging out-of-band RF frequencies during beamforming is to incorporate the {\em cross channel correlation} at mmWave bands with lower frequencies (2.4~GHz, radar bands, etc.). Such cross correlation is then utilized to reduce the beam search space by establishing a mapping between the channel measurements in the mmWave bands with lower frequencies (see Fig~\ref{fig:out-of-band}). Although the propagation characteristics in mmWave is different from lower frequencies, \textcolor{black}{recent research reveals that the main direction of arrivals (DoAs) are comparable.} Hence, the CSI at lower frequencies can be used to restrict the beam search space and avoid time-intensive exhaustive search, as proposed in the IEEE 802.11ad standard~\cite{yaman2016reducing}. This is relevant as mmWave systems are \textcolor{black}{very likely} to be deployed in conjunction with lower frequency systems, \textcolor{black}{where mmWave access points (APs) are envisioned to be paired with lower frequency APs that provide wide area control signalling and coordination. Moreover, multi-band communication is one of the proposed solutions for providing high throughput communication systems with high reliability, thus reinforcing the interest in taking advantage of such systems in the near future~\cite{kishiyama2013future}. Among the RF based out-of-band beamforming techniques, the use of radar signals and utilizing sub-6~GHz frequencies for mmWave beamforming have shown promising results.}   


\begin{table*}[hbtp]
    \setlength{\tabcolsep}{5pt}
    \centering
\resizebox{1.0\textwidth}{!}{
     \begin{tabular}{c|c|p{1.5cm}|p{1.5cm}|p{4cm}|p{1.5cm}|p{1.2cm}|p{4cm}}
    \specialrule{.2em}{.1em}{.1em} 
    \rowcolor{lightgray}
    {\bf Paper} & {\bf Year} & {\bf Transmission Frequency} & {\bf Dual Band Frequency} &{\bf Approach}& {\bf Validation} & {\bf Multiple AP} & {\bf Evaluation Metric} \\    
    \specialrule{.2em}{.1em}{.1em} 
    
    González-Prelcic {\em et al.}~\cite{gonzalez2016radar}  & 2016 & 76.5 GHz & 65 GHz & Compressive covariance estimation & Simulation & No & Sum spectral efficiency\\\hline
    Ali {\em et al.}~\cite{Ali2020PassiveLinks}  & 2020 &  73~GHz &  76~GHz & Compressive covariance estimation & Simulation & No & Transmission rate\\\hline
    Reus {\em et al.}~\cite{Reus2019BeamRadar}  & 2019 &  60~GHz &  5.89~GHz & Future location estimation & Simulation & No & Beamforming time\\\hline
    Nitsche {\em et al.}~\cite{Nitsche2015SteeringMeasurement}  & 2015 &  60~GHz &  2.4~GHz & LoS path detection & Experiment & No & Direct path estimation accuracy\\\hline
    Ali {\em et al.}~\cite{ali_2017}  & 2017 &  60~GHz &  3.5~GHz & DoA estimation & Simulation & No & Success percentage in detecting best beam\\\hline  
    Hashemi {\em et al.}~\cite{hasemi_2018}  & 2018 &  30~GHz &  3~GHz & DoA estimation & Experiment & No & mmWave throughput\\\hline    
    Mohamed {\em et al.}~\cite{Mohamed2015WiFiWLANs}  & 2015 &  60~GHz &  5~GHz & WiFi fingerprinting & Experiment & Yes & Average packet delay\\
      \specialrule{.2em}{.1em}{.1em} 
    \end{tabular}}
            \caption{Survey of literature on out-of-band RF based beamforming.}
    \label{tab:survey_out_of_band_RF}
    \vspace{-5mm}
\end{table*}

\vspace*{-0.2 in}
\subsection{Radar}
For a vehicle to infrastructure (V2I) hybrid MIMO scenario,  González-Prelcic {\em et al.}~\cite{gonzalez2016radar}, derives the channel information from \textcolor{black}{the infrastructure mounted radar} that is used to obtain  precoders/combiners at the vehicle and the infrastructure.
The radar sensor operates at 76.5~GHz, which is close to the mmWave communication band at 65~GHz. Taking advantage of this close proximity of the operating frequencies, the computed covariance of the received signal at the radar is applied as an estimation of the covariance of the communication signal in the mmWave band. The authors then argue that the optimum combiner is the dominant eigenvector of the covariance matrix of the received signal. Similarly, in the proposed scheme by Ali {\em et al.}~\cite{Ali2020PassiveLinks}, a passive radar at the \textcolor {black}{road side unit (RSU)} taps the radar signals transmitted by vehicle mounted automotive radars.
\textcolor{black}{In comparison to the prior works, the authors propose a simplified RSU based radar receiver that does not require the transmitted waveform as a reference for covariance estimation in ~\cite{ali2019spatial}.} To use the acquired radar information for mmWave beam initialization, a metric is defined that correlates the spatial information provided by the radar sensor and spatial characteristics of mmWave channel. This metric is then used to assess the the accuracy of the angular estimation.  Reus {\em et al.}~\cite{Reus2019BeamRadar} leverage the PHY layer \textcolor{black}{IEEE} 802.11ad frames to perform both radar operations and conventional communications using the standard compliant TX/RX chain. In this case, the radar is employed to estimate the location of vehicles, which is then used to select the \textcolor{black}{optimal mmWave beam.}
\subsection{Sub-6 GHz}
\textcolor{black}{Among the sub-6~GHz out-of-band beamforming techniques proposed in the state-of-the-art literature, Nitsche {\em et al.}~\cite{Nitsche2015SteeringMeasurement} propose a blind beam steering (BBS) system which couples mmWave with legacy 2.4/5 GHz bands to estimate} the direction for pairing nodes from passively overheard frames, as a replacement to the in-band trial-and-error beam initialization. Upon a beam training request, the proposed method first performs out-of-band direction inference to calculate angular profiles by broadcasting passively overheard detection band frames at the legacy sub-6~GHz band. In particular, an angular profile specifies received signal energy with respect to the azimuth incidence angle at the last known position. The LoS paths in all profiles remain nearly static, and appear as peaks at the same angle. However, the peaks resulting from reflections vary among profiles. Hence, the authors employ a profile history aggregation method over varying multipath conditions. By aggregating, the alternating reflection peaks are flattened and the remaining strongest peak \textcolor{black}{is estimated to correspond to the direct path}. Moreover, it gives an estimate of the uncertainty for the direct path estimate by measuring the deviation of the direct path angle over different profiles. Given the profile history for each device, a threshold for the \textcolor{black}{peak-to-average} ratio is defined to infer the LoS path and \textcolor{black}{ to reject the reflected paths.} 
If the ratio for a direction estimate is below this threshold, BBS proceeds with the legacy IEEE 802.11ad beam training method.
The experimental results depict that BBS successfully detects unobstructed direct path conditions with an accuracy of 96.5\% and reduces the IEEE 802.11ad \textcolor{black}{beam} training overhead by 81\%. 
\textcolor{black}{Similarly, in~\cite{ali_2017} the authors propose using the sub-6~GHz digital beam scanning method for faster estimation of the optimal direction. The candidate mmWave beams are restricted only to those beams that overlap with the dominant paths at sub-6~GHz band.} 
The angle of arrival (AoA) estimation on the 3~GHz channel is used in~\cite{hasemi_2018} to reduce the beam sweeping overhead for the mmWave in 30~GHz frequency. In particular, they experimentally show that in 94\% of LoS conditions, the identified AoA \textcolor{black}{in the 3~GHz band} is within $\pm 10 \degree$ accuracy for the AoA of the mmWave signal. Hence, the authors propose using MUltiple SIgnal Classification~(MUSIC) algorithm to estimate the AoA in the sub-6~GHz and running the exhaustive search only for angles 
\textcolor{black}{in the corresponding direction of the mmWave band, while factoring in the error bound of $\pm 10\degree$.} 

\textcolor{black}{One of the promising solutions for Gbps transmission in 5G is the use of wireless gigabit (WiGig) high frequency mmWave APs~\cite{wigigap}. However, multiple WiGig APs are required to fully cover the target environment, due to their short ranges. In this regard, a comprehensive network architecture along with a dual-band MAC protocol is proposed in~\cite{Mohamed2015WiFiWLANs} for coordinated WiGig WLANs, which is based on tight coordination between the 5~GHz (WiFi) and the 60~GHz (WiGig) unlicensed frequency bands~(see Fig.~\ref{fig:out-of-band}).} In the proposed dual-band MAC protocol operation, the control frames to be shared among the APs are transmitted via the wide coverage sub-6~GHz WiFi band, while the high speed data frames are concurrently transmitted by the WiGig APs in the mmWave band. \textcolor{black}{These control frames coordinate the beam training among the APs, so only one AP performs the beam training at a time, eliminating the probability of packet collisions due to simultaneous beamforming.} Also, the link information consisting of the used beam identification (ID), modulation coding scheme (MCS) index and received power, is broadcasted \textcolor{black}{in the sub-6~GHz WiFi frequencies, allowing other APs to} effectively exclude those beam IDs that may interfere with the existing data link from their beamforming training beams. 
\textcolor{black}{Moreover, since the location of a UE can be roughly estimated using WiFi channel information at WiFi frequencies through a process called fingerprinting, the authors propose this WiFi fingerprinting method to estimate the best and bad beam IDs of the WiGig links.}
Given a database of WiFi fingerprints and WiGig best beam IDs, an offline statistical learning is introduced where by comparing the current UE WiFi fingerprint with the pre-stored UE WiFi fingerprints, a best associated AP is selected for the UE, and a group of WiGig best sector IDs (beams) are estimated for the selected AP to effectively communicate with the UE at its current position. Among these estimated best beams, the beam IDs overlapping with the existing WiGig data links are recognized as bad beams and eliminated from the beamforming refinement process.

\textcolor{black}{We conclude the discussion on out-of-band RF based beamforming techniques by providing a comprehensive overview of these processes in Tab.~\ref{tab:survey_out_of_band_RF}. Next, we explore the existing challenges in this area.}

\subsection{Challenges}
\label{sec:RF_limitation}
While out-of-band RF assisted beamforming present promising improvements in beam initialization speed, there are some limitations associated, \textcolor{black}{which we itemize as follows: }
\begin{itemize}
    \item The out-of-band RF channel measurements need to be acquired constantly in order to estimate the channel at the mmWave band. Hence, it requires an integrated protocol for multi-band coexistence that can be challenging in dense networks.
    \item An optimal mapping is required between mmWave and out-of-band channel measurements. The mmWave band has unique propagation characteristics that preserves sparsity. In particular, the number of reflections is limited in mmWave band, while in lower frequencies, multiple reflections are normally observed. As a result, translating the DoA for bands that are located far apart from each other can be challenging and are prone to errors.
    \item RF-based out-of-band beamforming requires simultaneous multi-band channel measurements that increases the complexity of mmWave transceivers. Although future mmWave devices will likely support lower frequencies as well, this feature is not widely deployed in commercial devices yet.
    \item The existing out-of-band methods do not yet support simultaneous beamforming at both the transmitter and receiver sides, which is required for effective directional transmissions. 
\end{itemize}



\subsection{Non-RF Modalities for RF Tasks}
\label{sec:out-of-band-non-RF-for-RF-tasks}
Considering these challenges in out-of-band RF based beamforming techniques, 
there is growing interest in studying different non-RF modalities for optimizing wireless links. These various non-RF data modalities, e.g.,  RGB/RGB-D (RGB-Depth) camera images, LiDAR etc., capture the situational information in the environment from different perspectives, 
which can be exploited to assist in a variety of  wireless tasks, such as handover or channel quality prediction. While many sensing technologies have been proposed to enhance the reliability of wireless links, mostly in the mmWave bands,  recent advances in computer vision offer an untapped potential for  camera-aided communications. We discuss few examples of such applications and the proposed solutions, and present a comprehensive overview in Tab.~\ref{tab:survey_non_RF_task}. 


Most of the \textcolor{black}{existing research in using non-RF modalities} focus on problems that arise from the \textcolor{black}{unique} propagation characteristics, as well as susceptibility to blockage in the mmWave band. Oguma {\em et al.} in \cite{oguma2016proactive} propose a proactive mmWave base station selection method that predicts human blockage based on the dynamics observed through RGB-D camera images. Other works have taken similar approaches by exploring the use of camera images with reinforcement learning for handover management using single \cite{koda2019handover} and multiple cameras systems \cite{koda2020cooperative}. Jointly considering vision and communication is discussed in \cite{nishio2016high}, where Nishio {\em et al.} propose a network stack for a hybrid camera-communication system. 

\textcolor{black}{Recently there is also a rising interest in} predicting the channel quality without RF measurements. \textcolor{black}{Forecasting low SNR conditions or throughput reduction due to blockage or other channel metrics is the first step towards taking proactive measures before the link quality deteriorates, or worse, the connection is lost.} In particular, Nishio {\em et al.} in~\cite{nishio2019proactive} propose an RGB-D based received power prediction scheme for mmWave networks, based on multiple deep learning techniques to predict power losses up to hundreds of milliseconds ahead. Other works have analyzed specific challenges of this approach, such as the input data size~\cite{nakashima2018impact} or the application of pre-trained models in new scenarios using transfer learning \cite{mikuma2019transfer}. The fusion of in-band mmWave data with camera images is explored in \cite{koda2020communication} through a split-learning architecture, where the base station and the user run independent models and combine their predictions. Koda {\em et al.} show that a single pixel image can notably enhance the power prediction versus only-RF based approaches~\cite{koda2019one}. \textcolor{black}{In~\cite{Okamoto2017Machine-Learning-BasedCommunications}, a method to estimate throughput solely based on RGB-D images is presented by Okamoto {\em et al.}, with an RMS error of 114-178~Mbps in real time.} 

\begin{table*}[hbtp]
    \setlength{\tabcolsep}{5pt}
    \centering
\resizebox{0.9\textwidth}{!}{
     \begin{tabular}{c|p{3.6cm}|p{2.5cm}|p{4.7cm}|p{2.2cm}}
    \specialrule{.2em}{.1em}{.1em}  
    \rowcolor{lightgray}
    {\bf Paper} & {\bf Problem} & {\bf Data Type} &{\bf Approach}&  {\bf Evaluation Metric}\\    
    \specialrule{.2em}{.1em}{.1em} 
    
    Koda {\em et al.}~\cite{koda2019handover} & Handover  & RGB images & Reinforcement learning & Throughput\\\hline 
    Koda {\em et al.}~\cite{koda2020cooperative} & Handover  & RGB images & Reinforcement learning & Received power\\\hline
    Oguma {\em et al.}~\cite{oguma2016proactive} & Handover  & RGB-D images & Deterministic & Throughput\\\hline
    Nishio {\em et al.}~\cite{nishio2016high} & mmWave camera architecture  & Depth images & Network design & Throughput\\\hline
    Okamoto {\em et al.}~\cite{Okamoto2017Machine-Learning-BasedCommunications} & Throughput estimation  & RGB-D images & Adaptive regularization of weight vectors & Throughput\\\hline
    Oguma {\em et al.}~\cite{oguma2016performance} & Base station selection  & RGB-D images & Deterministic & Throughput\\\hline
    Mikuma {\em et al.}~\cite{mikuma2019transfer} & Received power prediction  & RGB-D images & Transfer learning & Received power\\\hline
    Nakashima {\em et al.}~\cite{nakashima2018impact} & Input data size for received power prediction  & Depth images & Convolutional LSTM & Input data size\\\hline
    Nishio {\em et al.}~\cite{nishio2019proactive} & Received power prediction  & RGB-D images & CNN, Conv. LSTM, and Random forest & Received power\\\hline
    Koda {\em et al.}~\cite{koda2020communication, koda2019one} & Received power prediction  & RGB-D images + RF & Split learning & Received power\\
      \specialrule{.2em}{.1em}{.1em} 

    \end{tabular}}
            \caption{Survey of literature leveraging the non-RF sensor data for solving different mmWave challenges.}
    \label{tab:survey_non_RF_task}
\end{table*}





After motivating the utility of leveraging various non-RF sensor data for RF tasks, we next map these benefits to the use-case of beamforming \textcolor{black}{in mmWave bands,} when higher magnitude of antenna elements (i.e., mMIMO systems) are involved. Additionally, the challenges of using RF-based out-of-band beamforming, described in Sec.~\ref{sec:RF_limitation}, suggest the research community needs to explore the space of  beyond RF-only solutions (be it traditional or RF-based out-of-band).
\textcolor{black}{We explore this direction in the next section.}

\section{Beamforming using Non-RF Sensor Data}
\label{sec:beamformingNonRF}
In mmWave beamforming, the location of the TX-RX pair and potential obstacles are the key factors that directly affect the optimal beam configuration.
Out-of-band RF aided beamforming methods estimate the approximate location of TX-RX pair given the AoA in  other RF bands, which helps to narrow down the beam search space. Interestingly, the situational state of the environment can also be acquired through data obtained from other  sensor devices~\cite{ali2020leveraging}, without occupying limited sub-6GHz RF resources.
This motivates the use of non-RF sensor data to speed up the beam initialization process in mmWave band~\cite{choi2016millimeter}.
Unlike the previously discussed out-of-band RF methods, the non-RF based beamforming does not require simultaneous multi-band channel measurements and optimal mapping between mmWave and CSI collected from another band. It is also capable of generating a mutually acceptable decision for both transmitter and receiver.

\begin{figure}
    \centering
    \includegraphics[width=0.8\linewidth]{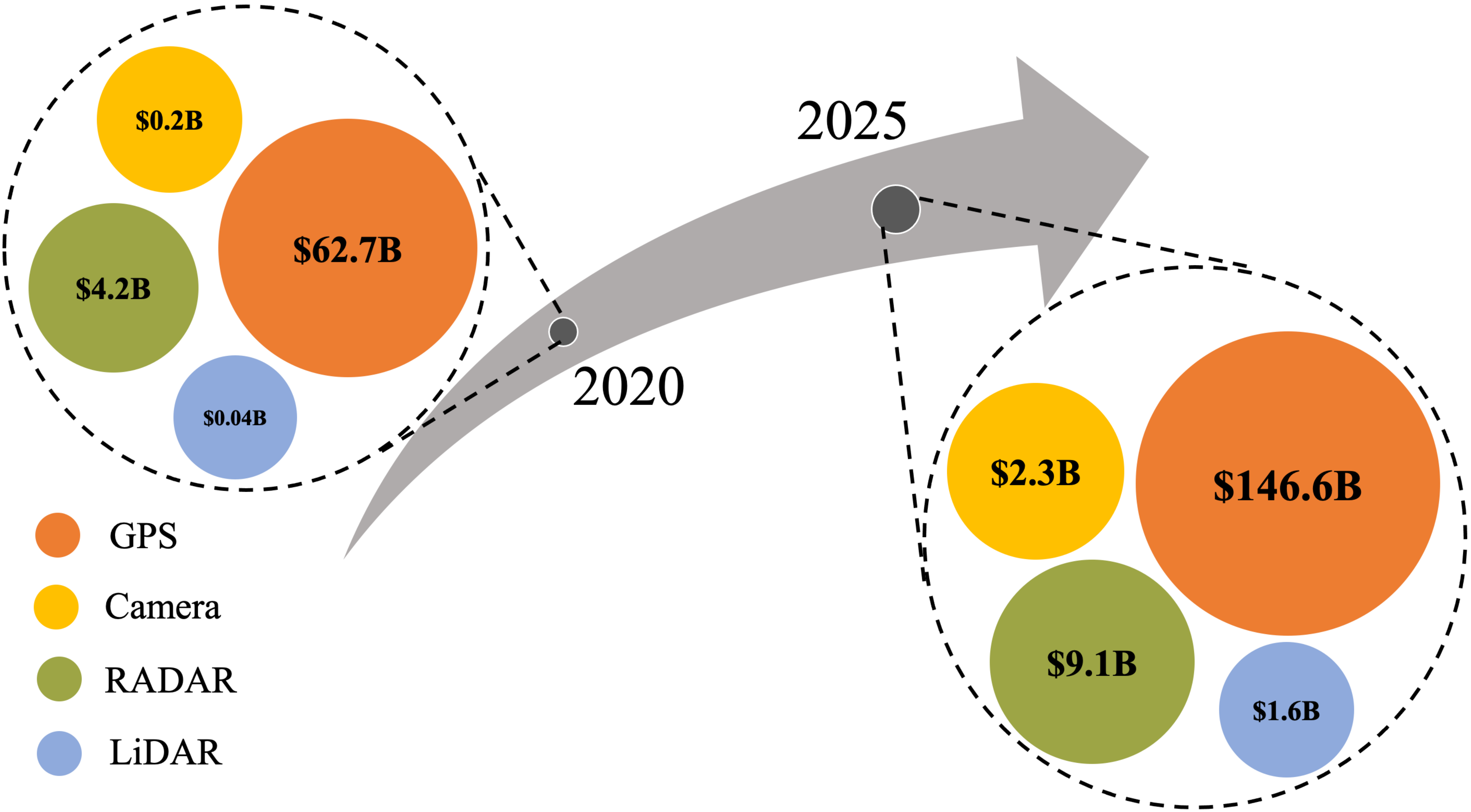}
    \caption{\textcolor{black}{Market revenue forecast for advanced driver assistance system (ADAS) sensors from the year 2020 to 2025.}}
    \label{fig:sensor_market}
    \vspace{-5mm}
\end{figure}

Typically, non-RF based beamforming utilizes inputs from a number of different sensors such as, GPS (Global Positioning System), camera, LiDAR (Light Detection and Ranging), which provide a 3-D representation of the surroundings, etc. \textcolor{black}{This is further aided by the fact that with the wide proliferation of IoT, multiple sensors are embedded in the environment, thus making it feasible to obtain  situational information from non-RF sources. As an example, consider the automotive sector with vehicles that have advanced driver-assistance systems~(ADAS).  Fig.~\ref{fig:sensor_market} depicts the increase in the market revenue of the various sensors enabling ADAS, as reported by Yole D\'evelopment~\cite{yole}.} It is expected that the global market for GPS, radar, cameras and LiDARs will reach $\$159.6$ billion in 2025. With the easy availability of such multitude of sensors, we need to incorporate methods  that leverage the heterogeneous sensor data to extract a rich understanding of the environment. 

In LoS scenarios, even though the optimal beam configuration can be estimated using the location of transmitter and receiver, it is not trivial to employ such approaches when encountering irregular radiation patterns, for e.g., when devices have multiple  side lobes. The problem becomes more challenging when estimating the strongest reflection from obstacles in NLoS conditions. Hence, a proactive method is required to learn the channel characteristics associated with the observed non-RF sensor modalities on a {\em case-by-case} basis. 
\textcolor{black}{Both deterministic and AI-enabled methods are proposed in literature that consider either single sensor modalities or multiple modalities through deep learning. We next go through these state-of-the-art methods, covering different sensor acquisition techniques, available datasets, exploitation methods of single and multiple modalities, and identify future research trends.}

\subsection{Data Acquisition and Processing}
\label{subsec:dataAcq_and_pre}
\textcolor{black}{
Choosing the right subset of sensor modalities to accurately capture the environment for detecting potential LoS paths and reflections affecting mmWave frequencies is crucial. The most popular sensor modalities for mmWave beamforming are presented below and their features are summarized in Tab.~\ref{tab:sensor_types}.}

\subsubsection{GPS} \textcolor{black}{This is a popular and widely available} satellite-based localization system that generates readings in the decimal degrees (DD) format, where the separation between each line of latitude or longitude (representing ${1}^{\circ}$ difference) is expressed as a float with 5 digit precision. Each measurement results in two numbers that together pinpoints the location on the earth’s surface. While localization accuracy in outdoor can be up to $2\, m$, it drastically decreases in indoor environment \textcolor{black}{because of GPS signal attenuation through walls and structures.} It is to be noted that the GPS sensor data refers to the latitude and longitude values generated from the GPS receiver, not the RF signals which are transmitted from the GPS satellites.

\subsubsection{\textcolor{black}{Image}} \textcolor{black}{Cameras can be used to} capture still RBG images of the environment and 
are commonly used in different applications such as cell phones and surveillance monitoring. Although images allow comprehensive environmental assessment, they are impacted by low-light conditions and obstructions.


\subsubsection{LiDAR} The Light Detection And Ranging sensor generates a 3-D representation of the environment by emitting pulsed laser beams. The distance of each individual object from the origin (i.e., the sensor location) is then calculated based on reflection times. The LiDAR can achieve much higher accuracy than image, but it is expensive and sensitive to weather conditions.



\begin{table}[t]
    \centering
     \begin{tabular}{c|c|c|c}
     \specialrule{.2em}{.1em}{.1em}  
     \rowcolor{lightgray}
     Sensing Technology & GPS & Camera & LiDAR \\ 
     \specialrule{.2em}{.1em}{.1em}  
     Cost & Low & Low &  High\\ \hline
     Horizontal Range & - & High &  High \\ \hline
     Resolution & Low & High &  High\\ \hline
     Night Vision & Yes & No &  Yes\\ \hline
     Affected by Weather & No & Yes &  Yes\\ 
     \specialrule{.2em}{.1em}{.1em}  
    \end{tabular}
            \caption{Sensing technologies for aiding the beamforming.}
    \label{tab:sensor_types}
    \vspace{-5mm}
\end{table}

\textcolor{black}{Even if a judicious choice is made on the sensor modality, simply using raw data} might fail to provide an accurate prediction. In particular, preprocessing on the raw data steps can improve the system performance many-folds as we describe later in this paper. 
Raw observations are not useful unless the role of each device that senses the data is specified, i.e. is the data captured from a transmitter, receiver or a potential obstacle? Each sensor type has it advantages and limitations. For example, GPS equipped objects can be utilized to track location, but these sensors cannot capture the presence of obstacles. LiDAR can collect the 3D state of the environment but fails to track the location of the target transceivers. Thus, GPS data can be merged with raw LiDAR data in the preprocessing step to mark the coordinates of the target receiver in the collected point clouds. Hence, data-level aggregation methods are one of the commonly used approaches to refine the the raw data to be more informative. Second, the preprocessing steps are also beneficial for reducing the data complexity by either discarding the irrelevant information or reducing the dimensionality of the input data. As an example, using a low-pass filter on camera images can reduce the dimensionality of the image by averaging the adjacent pixels while preserving the integrity. 
ML-based solutions only accept the data arranged in a fixed size, while for some modalities such as LiDAR the number of point clouds is varying on a case-by-case basis, depending on the number of present objects. Hence, preprocessing can account for this issue by transforming the data to a constrained representation without degrading information content. Hence, it is important to design proper preprocessing steps before using the data \textcolor{black}{for inference}. It should be noted here that the preprocessing pipeline of each modality must be designed based on the unique properties of each sensor type, while maintaining the highest correlation with the ultimate task, \textcolor{black}{which is mmWave beamforming in this case}.

\subsection{Datasets}
Next, we discuss the features of the available public datasets specific to beamforming using non-RF sensor modalities. These datasets enable the research community to explore different aspects of non-RF beamforming without incurring an individual effort of data collection.

\subsubsection{ViWi}
Alrabeiah {\em et al.} proposed a scalable synthetic framework called Vision-Wireless (ViWi)~\cite{alrabeiah2020viwi}. The scenario of interest is a V2I setting in 28~GHz mmWave band. The first release of this dataset contains four scenarios with different camera distributions (co-located and distributed) and views (blocked and direct). The channel characteristics and images are generated using the Remcom Wireless Insite ray-tracing~\cite{remcom} and Blender~\cite{blender} software, respectively. For each scenario, a set of images and raw wireless data (signal departure/arrival angles, path gains, and channel impulse responses) are recorded. An extended version of this dataset is  named ViWi vision-aided mmWave beam tracking (ViWi-BT)~\cite{Alrabeiah_viwidataset}, which contains 13 pairs of consecutive beam indices and corresponding street view images. This dataset contains a training set with 281,100 samples, a validation set with 120,468 samples, and a test set with 10,000 samples.

\subsubsection{Raymobtime}
\label{subsec:raymobtime_dataset}
The Raymobtime multimodal dataset~\cite{Klautau_2018} captures  \textcolor{black}{a virtual V2X deployment} with high fidelity in the urban canyon region of Rosslyn, Virginia for different traffic patterns. A static roadside BS is placed at a height of 4 meters, alongside moving buses, cars, and trucks. The traffic is generated using the Simulator for Urban MObility (SUMO) software~\cite{SUMO2018}, which allows flexibility in changing the vehicular movement patterns. The image and LiDAR sensor data are collected by Blender, and Blender Sensor Simulation (BlenSor)~\cite{blensor} software, respectively. For a so called scene, the framework designates one active receiver out of three possible vehicle types i.e. car, bus and truck. A python orchestrator invokes each software for each scene and collects synchronized samples of LiDAR point clouds, GPS coordinates and camera images mounted at the BS. The combined channel quality of different beam pairs are also generated using Wireless Insite ray-tracing~\cite{remcom} software. The number of codebook elements for BS and the receiver is 32 and 8, respectively, leading to 256 beam configurations overall. 


\subsubsection{Image-based} \textcolor{black}{This dataset is obtained by} Salehi {\em et al.} in~\cite{Salehi_2020} from a testbed composed of two Sibeam mmWave~\cite{ni_radios} antenna arrays mounted on sliders enabling horizontal movement. Using the mmWave transceivers from National Instruments, the mutual channel is measured for 13 beam directions at transmitter and receiver~(169 beam configurations overall). Two GoPro cameras observe the movements in the environment and are synchronized with the mmWave channel measurements. In the designed scheme, an obstacle blocks the LoS path between the transmitter and receiver and the experiment is repeated for two types of obstacles, wood and cardbox, causing 30dB and 4dB attenuation while blocking the LOS path, respectively.

\begin{table*}[hbtp]
    \setlength{\tabcolsep}{5pt}
    \centering
\resizebox{1.0\textwidth}{!}{
     \begin{tabular}{c|c|c|p{2.2cm}|p{2cm}|p{2cm}|p{1.3cm}|p{3.5cm}|p{2cm}}
    \specialrule{.2em}{.1em}{.1em}
    \rowcolor{lightgray}
    {\bf Paper} & {\bf Year} & {\bf Frequency} & {\bf Data Type} &{\bf Approach}& {\bf Validation} & {\bf Feedback Required?} & {\bf Evaluation Metric} & {\bf Dataset Publicly Available?}\\    
    \specialrule{.2em}{.1em}{.1em} 
    
    Kim {\em et al.}~\cite{kim2013enabling}  & 2013 & 60GHz & GPS & Deterministic & Simulation & No & Achievable capacity & No\\\hline
    
    Va {\em et al.}~\cite{va2016beam}  & 2016 & 60GHz  & GPS & Deterministic & Simulation & No & Outage and average rate ratio & No\\\hline
    
    Wang {\em et al.}~\cite{Wang_2018}  & 2018 & 5GHz & GPS & Deep learning & Simulation & No & Alignment probability & No\\\hline
    
    Va {\em et al.}~\cite{Va_2017}  & 2017 & 60GHz  & GPS & Deep learning & Simulation & Yes & Power loss probability & No\\\hline
    
    Sim {\em et al.}~\cite{sim2018online}  & 2018 & 28GHz & GPS & MAB & Simulation & Yes & Cumulative RX data & No\\\hline
     
    Aviles {\em et al.}~\cite{Aviles_2016}  & 2016 & 28GHz & GPS & Deterministic & Simulation & Yes & CDF of AoA estimation & No\\\hline

    Alrabeiah {\em et al.}~\cite{Alrabeiah_viwidataset}  & 2020 & 28GHz & Camera & Deep learning & Simulation & No & Top-1 accuracy & Yes\\\hline
    
    Tian {\em et al.}~\cite{tian2020applying}  & 2020 & 28GHz &  Camera & Deep learning & Simulation & No & Top-1 accuracy & Yes\\\hline
    
    Xu {\em et al.}~\cite{xu20203d}  & 2020 & 60GHz & Camera & Deep learning & Simulation & Yes & Top-K accuracy & Yes\\\hline
    
    Salehi {\em et al.}~\cite{Salehi_2020} & 2020 & 60GHz & Camera & Deep learning & Experiment & No & Top-1 accuracy & Yes\\\hline
    
    Woodford {\em et al.}~\cite{woodford2021spacebeam} & 2021 & {\color{black}28GHz} & LiDAR & Deterministic & Mixture & No & Link latency & No\\\hline

    Haider {\em et al.}~\cite{haider2018listeer} & 2018 & 60GHz &  Light sensor & Deterministic & Experiment  & No & CDF of AoA estimation & No\\\hline

    \rowcolor{lightcyan}
    Klautau {\em et al.}~\cite{klautau_2019} & 2019 & 60GHz & GPS and LiDAR & Deep learning & Simulation & Yes & Top-K accuracy & Yes\\\hline
    
    \rowcolor{lightcyan}
    Dias {\em et al.}~\cite{dias2019position} & 2019 & 60GHz &  GPS and LiDAR & Deep learning & Simulation & Yes & Top-K accuracy & Yes\\\hline
    
    \rowcolor{lightcyan}
    Alrabeiah {\em et al.}~\cite{alrabeiah_image_sub6} & 2020 & 28GHz & Camera and sub-6 & Deep learning & Simulation & Yes & Top-K accuracy & Yes\\
    
      \specialrule{.2em}{.1em}{.1em} 

    \end{tabular}}
    \caption{Survey of literature on non-RF data for out-of-band and multimodal beamforming. The highlighted three rows of the table depict the most recent effort of using multimodal non-RF sensor data to aid in beamforming, referred as {\em multimodal beamforming} in this article.}
    \label{tab:survey_non_RF}
    \vspace{-5mm}
\end{table*}

\subsection{Single non-RF Modalities}
\label{subsec:single_non_RF}

Next, we present detailed descriptions of different studies and algorithms that use a single non-RF sensor modality. These include either GPS coordinates, camera images or LiDAR point clouds to accelerate the beam selection, and by extension, the beamforming process. Multimodal fusion is described later in Sec.~\ref{subsec:multimodal_beamforming}.

\noindent
$\bullet${\bf GPS Coordinates:} The knowledge of the location of target receiver has been used earlier to address the challenges of cell discovery~\cite{capone2015obstacle}. The same idea can be used to speed up the beam initialization in mmWave band, \textcolor{black}{which utilizes directional transmission}. The authors in~\cite{kim2013enabling,va2016beam} use the \textcolor{black}{GPS based} position of the receiver to estimate the optimum future beam directions. In particular, the proposed algorithms predict the future locations by tracking the mobility profile of the receiver and geometrical features of the environment. However, it should be noted that this approach only works when the LoS path is available. 
Alternatively, Wang {\em et al.} propose a framework for mmWave beam prediction by exploiting the situational awareness~\cite{Wang_2018}. They use the location of all the vehicles in the same scene as features to extend the solution to NLoS scenarios. The simulation scenario consists of small cars and trucks, any of which can be the target receiver. The authors \textcolor{black}{argue that} the vehicle dynamics have the main effect on the optimum beam configuration, since the road side buildings and infrastructures are stationary, and pedestrians are small in size. Hence, a feature vector map $v = [r,t_1,t_2,c_1,c_2]$ is generated where $r$ depicts the location of RSU, $t$ and $c$ represent the truck and car vehicles. The subscripts 1 and 2 denote the lane index where the vehicle is located and each vector $(t_i,c_i),~i=1,2$ includes the location of the corresponding vehicle type in ascending order for the lane $i$. Since the ML algorithms accept a fixed size input, the number of trucks/cars on each lane is constrained, and the vehicles which are far away are eliminated. This feature vector is then used to predict the received power for any beam in the codebook, by leveraging ML. 
Similarly, Va {\em et al.}~\cite{Va_2017} propose an algorithm where the location of all the vehicles on the road, including the target receiver, is used as input to an ML algorithm, to infer the best beam configuration. The proposed algorithm uses the power loss probability as a metric to estimate the misalignment probability that might occur \textcolor{black}{when non-optimal beams are selected}. \textcolor{black}{In this case,} a subset of the beam configurations are suggested \textcolor{black}{by the authors} to minimize this misalignment probability. 
In order to speed up the beam initialization, an online learning algorithm is proposed in \cite{sim2018online}, which exploits the coarse user location information in vehicular systems. In particular, the problem is modeled as a contextual multi armed bandit (MAB) problem and a lightweight context-aware online learning algorithm, namely fast machine learning (FML) is used to learn from and adapt to the environment. The proposed FML algorithm explores different beams over time while accounting for contextual information (i.e., vehicles’ direction of arrival) and adapts the future beams accordingly, \textcolor{black}{in order} to account for the system dynamics such as the appearance of blockages and changes in traffic patterns. In comparison, Aviles {\em et al.} in~\cite{Aviles_2016} first generate a database that captures the propagation characteristics at 28~GHz and the position of UE. 
\textcolor{black}{Then, given the location of a UE, a hierarchical alignment scheme is proposed, which consults with this database and incorporates the position of the UE for faster beam alignment.}


\noindent
$\bullet${\bf Camera Images and Light Sensors:} The cameras are \textcolor{black}{one of the sensing modalities that  capture} the situational state of the environment with high resolution. With the recent progress in computer vision and deep learning, powerful algorithms are now available that can be used for processing the images \textcolor{black}{in real time} for beamforming. A baseline for ViWi-BT dataset is presented in \cite{Alrabeiah_viwidataset} based on gated recurrent units (GRUs) without the images and only the sequence of beam indices. The authors argue that beam prediction accuracy is expected to improve significantly by leveraging both wireless and visual data. In \cite{tian2020applying}, Tian {\em et al.} propose a framework to predict future beam indices from previously observed beam indices and images. The proposed approach consists of three steps as follows. The first step consists of feature extraction, where ResNet, ResNext and 3D ResNext modules, \textcolor{black}{each proven to have powerful feature-representation abilities}, are used to capture 2D and 3D spatio-temporal features from the images. 
In the second step, a long short-term memory (LSTM)~\cite{hochreiter1997long} network is designed to incorporate the time-series data for prediction. Finally, a feature-fusion module aggregates features from ResNet and 3D ResNext to generate high-level features. The fusion module comprises of two LSTM networks and a simple cross-gating block that only support linear transformation. To validate their approach, the authors use ViWi-BT dataset where the first eight pairs of images are used to predict next five future beams.
Similarly, in \cite{xu20203d}, Xu {\em et al.} propose a scheme where the images captured from different perspectives are used to construct a 3D scene that resembles the point cloud data collected by 3D sensors like LiDAR. Then, a CNN with 3D input is designed to predict the future beams to be selected. Results rreveal that the proposed 3D scene based beam selection outperforms LiDAR in  accuracy, without imposing the huge cost of LiDAR sensor. While the majority of current literature uses synthetic datasets, the authors in~\cite{Salehi_2020} deploy a testbed using National Instruments radio at 60 GHz~\cite{ni_radios} and camera generated images to predict the best beam configuration. Their proposed method consists of two main steps, namely detection and prediction. In the first step, the transmitter and receiver are detected in the image in the form of a bitmap. This step is important \textcolor{black}{to detect the features which are relevant to the task} and discard the irrelevant ones, such as static walls, etc. Finally, the bitmaps are fed to another CNN to predict the optimum beam configuration given the historical data from collected dataset.
The LiSteer system proposed in~\cite{haider2018listeer} steers mmWave beams to mobile devices by re-purposing indicator light emitting diodes (LEDs) on wireless APs to passively track the direction to the AP using light intensity measurements with off-the-shelf light sensors. The proposed approach considers the pseudo-optical properties of mmWave signal, i.e., dominant LoS propagation, to approximate the APs' AoA in the mmWave band. Hence, their approach requires the APs to be equipped with LEDs \textcolor{black}{and to be situated} close to the mmWave band antenna. The authors propose using an array of light sensors to combat the in-coherency of light-AoA estimation that also allows steering beams for both 2D and 3D beamforming codebooks. The experimental results demonstrate that LiSteer achieves direction estimates within $2.5\degree$ of ground truth on average with beam steering accuracy of more than 97\% in tracking mode, without incurring any client beam training or feedback overhead.  

\noindent
$\bullet${\bf LiDAR Point Clouds:} Woodford {\em et al.}~\cite{woodford2021spacebeam} use LiDAR to build a 3D map of the surrounding physical environment and captures the characteristics of the physical materials. The proposed approach uses a customized ray-tracing algorithm that can identify real RF paths in a 3D mesh generated by LiDAR sensors, and reject false reflection paths caused by reconstruction noise. The output of this phase is a pre-computed look-up table to select the best beams for all mmWave links in the environment. It should be noted that the LiDAR sensors are not required during the ordinary operation of the system and are only used in advance to generate the lookup table. The proposed approach can recompute the complete lookup table for the environment within 15 minutes. The authors validate their approach using  Azure Kinect LiDAR camera~\cite{azure} and a commercial 802.11ad radio~\cite{airfidenet}, 
yielding to 66\% reduction in latency and 50\% increase in throughput.

\subsection{Multimodal Beamforming}
\label{subsec:multimodal_beamforming}
Since, each of the above sensor modalities capture different aspects of the environment, using more than one sensor modality and intelligently fusing these multimodal data can result in more comprehensive understanding of the environment and can consequently enable the undertaking of robust decisions.

\noindent
$\bullet$ {\bf Benefits of Fusion:} 
\textcolor{black}{The fusion of multimodal data over the single modalities has multiple advantages, as explained below:}
\begin{itemize}
    \item \textit{Enhanced Data Representation:} For the situational information to be effective during beamforming, it is crucial to differentiate between the transmitter, receiver and obstacles. However, some sensor modalities cannot provide such information by only relying on raw data. In this case, the data from different modalities can be \textcolor{black}{fused together} to improve the data representation. As an instance, it is not trivial to locate the receiver within a  LiDAR point cloud. In this case, the GPS coordinates can be used to mark the target receiver.
    \item \textit{Compensate for the Missing Information:} Sometimes the captured data from each sensing modality reflect an aspect of the \textcolor{black}{environment}, yet none can provide a complete understanding by it's own. For instance the dimensionality of objects is not reflected in GPS, and the accurate Cartesian coordinates of the target receiver cannot be acquired using LiDAR or image sensors.
    \item \textit{\textcolor{black}{Improved Accuracy:}} Using more than one modality \textcolor{black}{enables a fine grained understanding of the environment} which results in more accurate predictions. Hence, fusion reinforces the prediction accuracy by gathering the information from different sensors to make the final decision. In this case, the fusion algorithms can automatically adjust the weights of each modality towards the optimum performance.
    \item \textit{Robustness to Errors:} Collecting data using sensor devices come with associated considerations, including the inherent error. Here, the accuracy of measurement is dependent on working with the nominal structure that the device is designed for. For instance, the accuracy of LiDAR sensor degrades in with sunlight reflections, while it does not affect the GPS data~\cite{heinzler2019weather}. Hence, fusion increases prediction robustness in the case of inaccurate or unreliable data. 
    \item \textit{Availability:} In some applications, the sensor \textcolor{black}{does not have to} be co-located. Hence, secondary control channels are required to enable the connectivity between the different sensors and the computing unit. However, this control channel is also subject to saturation and loss. Using more than one modality with fusion helps the system to be robust to such scenarios and it guarantees that the prediction happens when at least one modality is available during the inference. 
\end{itemize}
\textcolor{black}{Below, we give some examples of state-of-the-art multimodal beamforming with  different fusion approaches on multiple sensors.}\\
\noindent
$\bullet$ {\bf GPS and LiDAR Fusion:} \textcolor{black}{Consider a typical V2I setting, where a static BS wants to establish communication with a target vehicle-mounted receiver. The vehicle is assumed to be equipped with GPS and LiDAR sensors} that enable the vehicle to acquire its location and detect blocking objects nearby. In this scenario, Klautau {\em et al.} propose a distributed architecture to reduce the mmWave beam selection overhead~\cite{klautau_2019}. Here, the BS constantly broadcasts its position via a low-band control channel. The situational state of the environment is then collected using LiDAR, situated on the vehicle and is aggregated by BS location in the preprocessing pipeline, \textcolor{black}{where a histogram is generated at the beginning} to quantize the space. The LiDAR point clouds then lie in the corresponding bin of the histogram, and the location of BS and receiver is also marked with unique indicators. Using the proposed preprocessing step, the measured point clouds are mapped to a ridge \textcolor{black}{represented by a fixed size}. Note that the number of point clouds in the raw data \textcolor{black}{varies} depending on the number of \textcolor{black}{objects present} during the measurement. This refined data representation is then fed as input to a deep CNN to estimate a set of $K$ most likely candidate beam pairs. The selected beam pairs are then sent to the BS, \textcolor{black}{and beam training is performed to generate the suggested subset} to obtain the optimum beam configuration. Similarly, \cite{dias2019position} considers a V2I setting and compares the performance of the previously described distributed scheme with two centralized schemes: \textcolor{black}{(i)} using a single LiDAR located at the BS, and \textcolor{black}{(ii)} fusing LIDAR data from neighboring vehicles at the BS. The LiDAR data is then used for both LOS detection and beam selection for three competing scenarios. The experimental results in this work depicts that in LoS, distributed and centralized methods perform closely, while the LiDAR at BS results in lower top-$K$ beam prediction accuracy, because of limited range of LiDAR. On the other hand, in NLoS scenarios, the distributed scheme outperforms the centralized method, and both are better than LiDAR at BS. 

\noindent
$\bullet$ {\bf Camera with Sub-6 GHz Fusion:} The possibility of vision-aided wireless communications is evaluated in~\cite{alrabeiah_image_sub6} where a camera at the BS observes the movements in the environment, and snapshots of the environment are paired with sub-6 GHz channels to help overcome the beam selection and blockage prediction overhead. The proposed method models the beam prediction from images as an image classification task. Hence, each user location in the scene is mapped to a class representing the associated beamforming codebook. However, the pure image input may be insufficient for blockage detection since the instances of `no user' and `blocked user' are visually the same. Hence, in order to identify blocked users, the images are fused by sub-6~GHz channels to account for the aforementioned challenge. 

The concise overview of different state-of-the-art beamforming methods while using single or multimodal data is presented in Tab.~\ref{tab:survey_non_RF}.



\subsection{Future Research Directions}
While multimodal learning is an extremely interesting research field, there are some challenges that need to be addressed. First, in order to exploit more than one modality, the synchronized information of all modalities must be present during inference. This requires a precise network controller and back-channel to enable connectivity among different modules while accounting for privacy concerns. Second, the fusion scheme needs to be designed such that the different modalities result in a reinforced prediction. The fusion model can be as simple as a linear transformation, such as summation or multiplication. However, learning the relation between different modalities might require non-linear transformations such as deep learning on custom-made neural networks.
\textcolor{black}{In that regard, we explore few novel fusion techniques that use non-linear transformation in the following section.}

\section{Detailed Description of Data Fusion Methods for Multimodal Beamforming}
\label{sec:proposed_work}

\begin{figure*}[t!]
\centering
\includegraphics[width=0.8\textwidth, trim=2cm 2cm 2cm 1cm, clip]{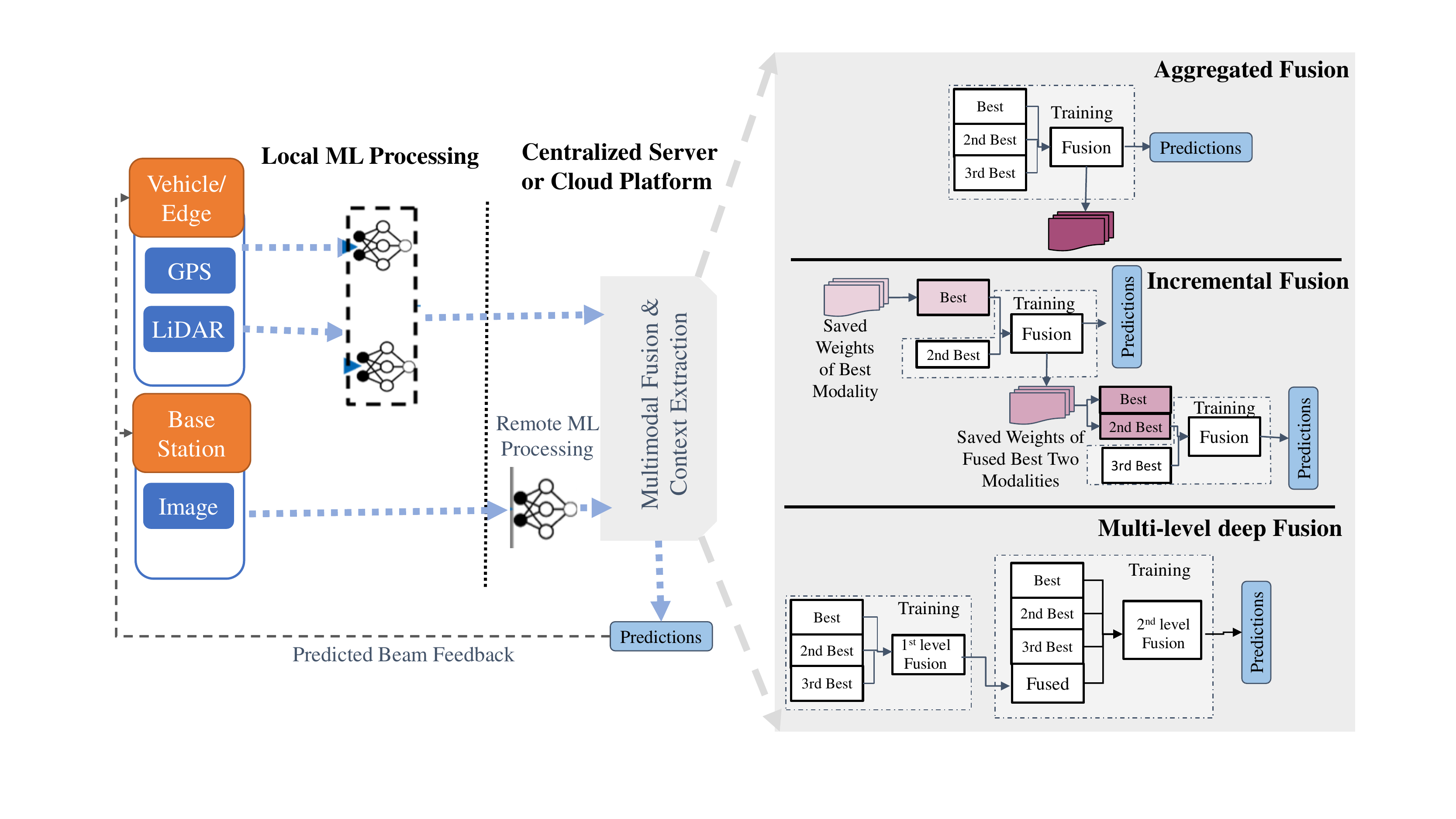}
\caption{A snapshot of our proposed multimodal beamforming pipeline with novel fusion techniques at a vehicular network. Different sensor modalities are captured in the vehicle or base stations and run through few local ML feature extraction phase. The context extraction and multimodal fusion can therefore be deployed in the connected centralized system or the cloud platform. The prediction from the centralized system or cloud is fed back to the vehicle and the base station for a successful beamforming scenario.}
\label{fig:scenario_of_interest}
\vspace{-5mm}
\end{figure*}

In the earlier sections, we make the case for using non-RF modalities for mmWave beamforming, and more specifically, the motivation for using multimodal data to capture the holistic information of the wireless environment. Successfully exploiting multiple modalities depends on the ability of {\em how} and {\em when} to fuse different modalities. The state-of-the-art deep learning based fusion techniques generally involve concatenation of layers \cite{Perez-Rua_2019, Chadwick_2019, Chen_2017_CVPR}, element wise operation~\cite{Nobis_2019, Chen_2017_CVPR} and cross-modality analysis~\cite{Liu_2020, Lin_2020} in an one-step manner. In other words, all the available modalities are generally fused once in an aggregated manner. One of the contributions of this survey is to explore the full potential of multimodal learning by analyzing diverse modality fusion techniques for beam selection algorithms.

 Novel fusion methods based on deep learning flexibly assign weights on each modality based on their relevance from the situational state information. This results in a faster yet accurate beam selection solution. Our vision for multimodal beamforming is to generate novel {\em feature-level} and {\em multi-level} fusion frameworks for fast and accurate beamforming by minimizing the beam search space. By fusing the features/outputs from various {\em unimodal models}, we describe a platform to improve the individual prediction accuracy. The customized neural network architectures for each single modalities are referred as unimodal models in this paper. We propose fusion frameworks which work on the {\em ultimate} and {\em penultimate} layers of each unimodal model.
The multi-level aspect of the proposed framework is as follows: a feature-based fusion takes place at the first level, followed by subsequent fusions at each level based on the prediction of the previous one. 

\subsection{Scenario of Interest}
As before, consider a vehicular communication scenario need for  autonomous cars, where seamless yet ultra-fast communication between moving vehicles is crucial. Multiple sensors are already included as standard installations for the majority of new vehicles, as well as in fixed roadside base stations~\cite{Gonzalez_2017}. For example, LiDAR sensors are an indispensable part of modern vehicles that are used for either automated driving or collision avoidance~\cite{Premebida_2007}. The GPS data are regularly collected and transmitted as part of basic safety messages frame in V2X applications~\cite{Festag_2015}, and surveillance cameras have been used for decades as part of smart-city initiatives ~\cite{Liu_2016}. 
The V2X network scenario using these three sensor data types and exploiting various fusion techniques for beamforming is illustrated in Fig.~\ref{fig:scenario_of_interest}. Next, we formulate the problem of reduced space beam selection which leverages the non-RF multimodal data in the multimodal beamforming problem. 

\subsection{Formulation of Top-$K$ Beam Selection for Multimodal Beamforming}
\label{subsec:problem_formulation}
We consider a codebook-based mMIMO transmitter and receiver, where each codebook element represents a particular beam direction. The transmitter and receiver codebooks are represented as: $C_{tx}=\{t_1,\dots,t_M\},~C_{rx}=\{r_1,\dots,r_N\}$, where $M,N$ are the number of transmitter and receiver codebook elements, respectively.
Hence, the set of all possible beam pairs $\mathcal{B} = \{(t_{m},r_{n})|t_{m}\in C_{tx},r_{n}\in C_{rx}\}$,
with $|\mathcal{B}|=M\times N $. For a specific beam pair $(t_m,r_n)$, the normalized signal power is $p_{(t_m,r_n)} = |w_{t_m}^* ~\mathbb{H}~w_{r_n}|^2$, where $\mathbb{H}\in \mathbb{R}^{M\times N}$ is the channel matrix  and $*$ is the conjugate transpose operator. The weights $w_{t_m}$ and $w_{r_n}$ indicate the corresponding beam weight vectors associated with the codebook element $t_m$ and $r_n$, respectively ($|w_{t_m}|=M,|w_{r_n}|=N$). The set of normalized power for all beam configuration is defined by: $\mathcal{P} = \{p_{(t_m,r_n)}|t_{m}\in C_{tx},r_{n}\in C_{rx}\}$. The beam selection process for multimodal beamforming is defined to find top-$K$ best beam configuration, $\mathcal{B}_K$:
\begin{equation}
    \mathcal{B}_K = 
     \underset{A\subseteq \mathcal{B}, |A|=K}{\arg\max}~A\in\mathcal{P}.
    \label{eq:argmax_q}
\end{equation}
The classical standards (both IEEE 802.11ad~\cite{Nitsche2014IEEEWi-Fi} and 5G-NR \cite{Giordani2019StandaloneMmwaves}) for beam selection corresponds to sweeping all beam pairs $(t_m,r_n) \in \mathcal{B}$ sequentially in order to find the best one for beamforming. The selected set of $\mathcal{B}_K$ restricts the search for the optimal pair to this set. 

Recall that the multimodal data to be comprised of GPS, LiDAR and camera image sensor data. The dimensionality of the data matrices are defined as: $(d_0^\ldr \times d_1^\ldr \times d_2^\ldr)$ and $(d_0^\img \times d_1^\img)$ for 3D LiDAR and 2D image sensors, respectively. GPS coordinate has 2 elements: latitude and longitude. Next we define each unimodal data separately as: $X_{\ldr}\in \mathbb{R}^{\mathcal{N}\times d_0^\ldr \times d_1^\ldr \times d_2^\ldr}, X_{\img}\in \mathbb{R}^{\mathcal{N}\times d_0^\img \times d_1^\img},  X_{\crd}\in \mathbb{R}^{\mathcal{N}\times 2}$, for LiDAR, image, and coordinate, respectively, where $\mathcal{N}$ is the number of training samples. The overall multimodal data is represented as: $X = [X_{\ldr};~X_{\img};~X_{\crd}]$. We configure the label matrix $Y\in \{0,1\}^{\mathcal{N}\times\mathcal{|B|}}$ to represent the basis vector of $\mathcal{B}$ beam pairs, where the optimum beam pair is set to $1$, and rest are $0$, following the Eq. \eqref{eq:argmax_q}. Next, we define different unimodal models and the necessary notations used for proposed fusion frameworks.

\subsection{Unimodal Models}
Each unimodal network works as a feature extractor for each modality. The details of each {\em unimodal network} are out of the scope of this survey. However, modeling the penultimate and ultimate layers of each candidate modality is necessary for formulating different fusion approaches. The latent embeddings from each unimodal feature extractor are captured at the penultimate layer, and the prediction from each unimodal network is captured at its ultimate layer.

\noindent
$\bullet$ \textbf{Defining the Latent Embeddings:}
We assume that the penultimate layers of the unimodal networks of LiDAR, image, and GPS coordinates have $d^\ldr$, $d^\img$, $d^\crd$ neurons, respectively. Hence, sample space of the inputs of each sensor modality maps to a vector with a dimension equal to corresponding number of neurons.

We denote the feature extractor of each modality as $f_{\theta^\ldr}^\ldr$, $f_{\theta^\img}^\img$, and  $f_{\theta^\crd}^\crd$ for LiDAR, image, and coordinate data, respectively, each parameterized by weight vectors $\theta^m$, for $m\in\{\ldr,\img,\crd\}$. We refer to the output of these feature extractors as the latent embedding of each modality. Formally,%
\begin{subequations}%
    \begin{align}
        & \mathbf{z}_{\ldr} = f_{\theta^\ldr}^\ldr(X_{\ldr}),~~~~~~f_{\theta^\ldr}^\ldr:\mathbb{R}^{d_0^\ldr\times d_1^\ldr} \mapsto \mathbb{R}^{d^{\ldr}}\\
     & \mathbf{z}_{\img} = f_{\theta^\img}^\img(X_{\img}),~~~~~~f_{\theta^\img}^\img:\mathbb{R}^{d_0^\img\times d_1^\img} \mapsto \mathbb{R}^{d^{\img}} \\
     & \mathbf{z}_{\crd} = f_{\theta^\crd}^\crd(X_{\crd}),~~~~~~f_{\theta^\crd}^\crd:\mathbb{R}^{d_0^\crd\times d_1^\crd\times d_2^\crd} \mapsto \mathbb{R}^{d^{\crd}}
    \end{align}
\label{eq:feature_extrs}%
\end{subequations}
where $\mathbf{z}_{\ldr}$, $\mathbf{z}_{\img}$, and $\mathbf{z}_{\crd}$ show the extracted latent embeddings for input data $X_{\ldr}$, $X_{\img}$, and $X_{\crd}$ respectively.

\noindent
$\bullet$ \textbf{Defining the Ultimate Layers:}
We formulate the output of the ultimate layers for each unimodal network with regard to the corresponding latent embeddings. The ultimate layers are just the transformation the penultimate embeddings with suitable activation functions. The output of ultimate layers for unimodal networks of LiDAR, image, and coordinate are denoted as $\scores_{\ldr}$, $\scores_{\img}$, and $\scores_{\crd}$, respectively,
\begin{subequations}%
    \begin{align}
        & \scores_{\ldr} = \sigma(f_{\theta^\ldr}^\ldr(\mathbf{z}_{\ldr})),~~~~~f_{\theta^\ldr}^\ldr:\mathbb{R}^{d_0^\ldr\times d_1^\ldr} \mapsto \mathbb{R}^{|\mathcal{B}|} \\
     & \scores_{\img} = \sigma(f_{\theta^\img}^\img(\mathbf{z}_{\img})),~~~~~f_{\theta^\img}^\img:\mathbb{R}^{d_0^\img\times d_1^\img} \mapsto \mathbb{R}^{|\mathcal{B}|} \\
     & \scores_{\crd} = \sigma (f_{\theta^\crd}^\crd(\mathbf{z}_{\crd})),~~~~~f_{\theta^\crd}^\crd:\mathbb{R}^{d_0^\crd\times d_1^\crd\times d_2^\crd} \mapsto \mathbb{R}^{|\mathcal{B}|} 
     \vspace{-2mm}
    \end{align}
\end{subequations}
where $\sigma$ is the softmax activation functions over the latent embeddings. Finally, the output of each ultimate layer maps to all possible beam pairs $\mathcal{B}$ combining transmitter and receiver codebook elements.
\vspace{-5mm}

\subsection{Novel Feature-level Fusions at Penultimate Layer}
\subsubsection{Aggregated Fusion}
\begin{figure}[t!]
\centering
\includegraphics[width=0.45\textwidth]{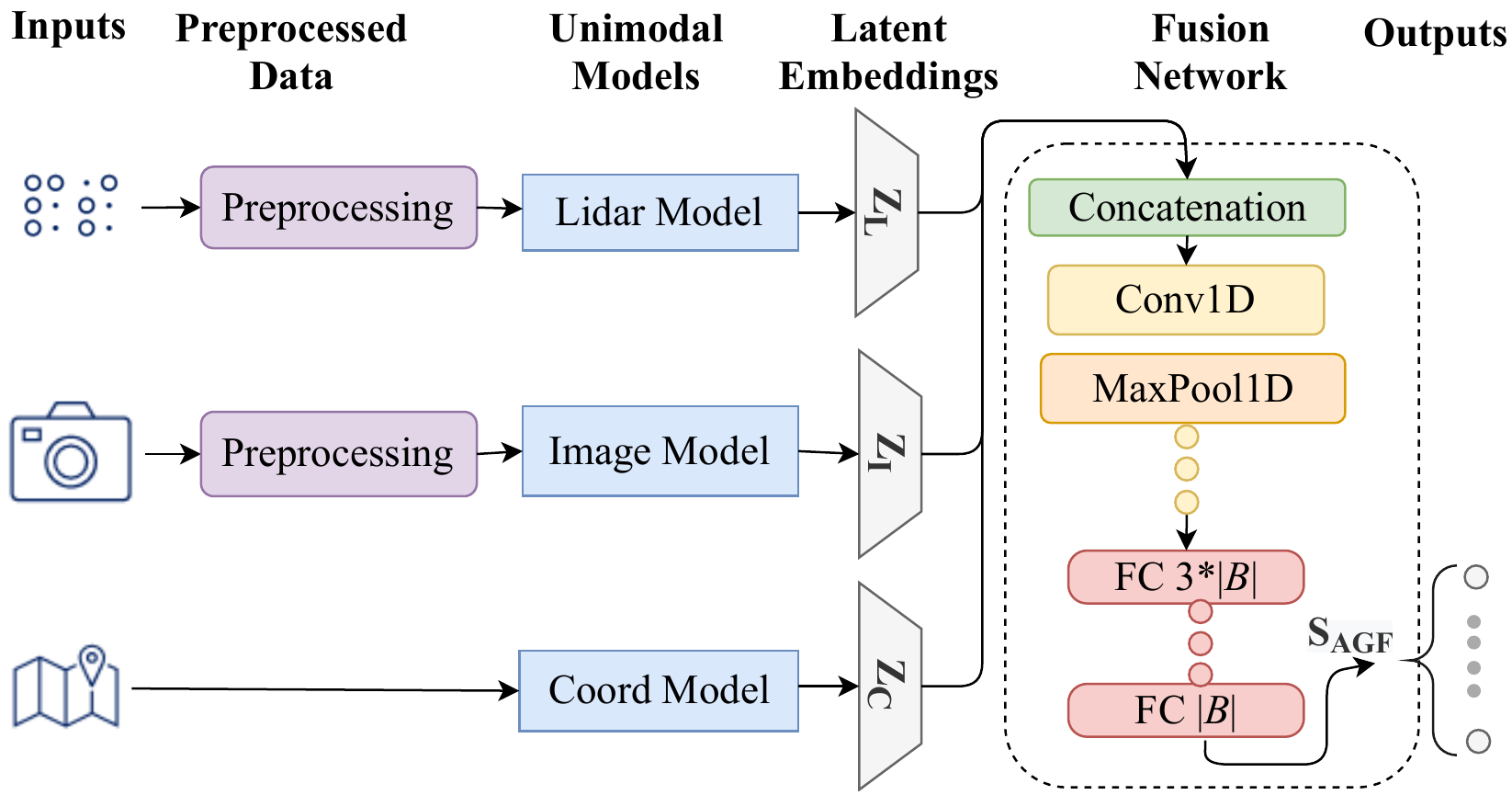}
\caption{Proposed aggregated fusion at penultimate layer for multimodal beamforming. The output of fusion maps to all possible beam pairs $\mathcal{B}$.}
\label{fig_fusion_penul}
\end{figure}

The conceptual overview of the aggregated fusion network is shown in Fig.~\ref{fig_fusion_penul}. The latent embeddings of the penultimate layers from each unimodal network are concatenated in an aggregated manner. The fusion network can be designed with multiple convolutional, pooling and fully connected layers afterwards. 
The overall concept is generalized as follows: first, given $\mathbf{z}_{\ldr} \in \mathbb{R}^{d^{\ldr}}$, $\mathbf{z}_{\img}\in \mathbb{R}^{d^{\img}}$ and $\mathbf{z}_{\crd} \in \mathbb{R}^{d^{\crd}}$, we concatenate them in aggregated manner and generate the combined latent feature matrix $\mathbf{z}$:
\[
    \mathbf{z} = [\mathbf{z}_{\ldr} ; \mathbf{z}_{\img}; \mathbf{z}_{\crd}]\in \mathbb{R}^{d^{\ldr}+d^{\img}+d^{\crd}}.\vspace{-2mm}
\]
Them, we denote the aggregated fusion network as $f_{\theta^\AGF}^\AGF(.)$. Finally, we use softmax activation function ($\sigma$) to predict the optimality of each beam as:
\[
\vspace{-2mm}
 \scores_{\AGF} = \sigma (f_{\theta^\AGF}^\AGF(\mathbf{z})),~~~~~f_{\theta^\AGF}^\AGF:\mathbb{R}^{d^{\ldr}+d^{\img}+d^{\crd}} \mapsto \mathbb{R}^{|\mathcal{B}|}
\]
where $\scores_{\AGF}$ is the ultimate layer output of the aggregated fusion network.

\begin{figure}[t!]
\centering
\includegraphics[width=0.5\textwidth, trim=0.5cm 4cm 10cm 0cm, clip]{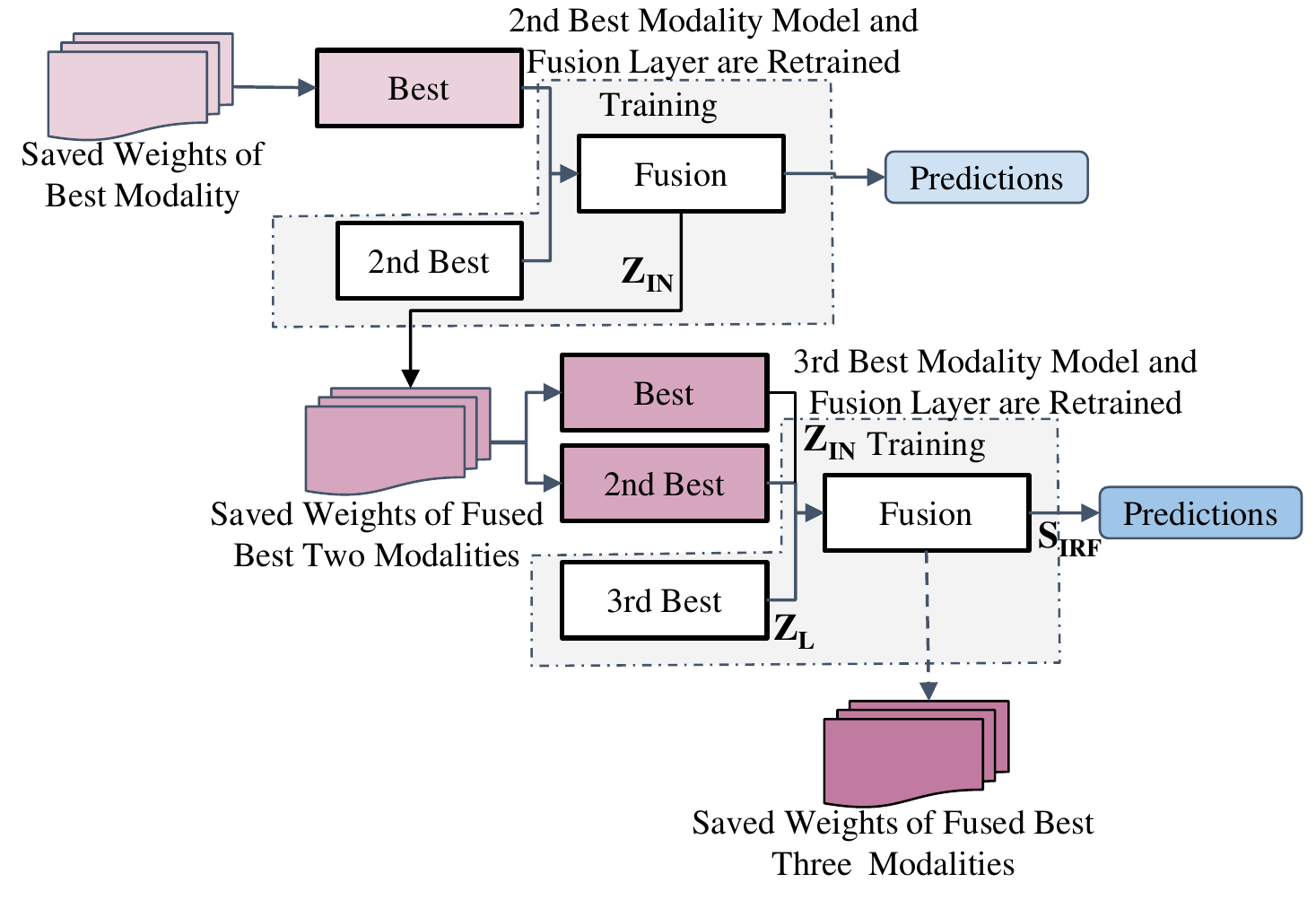}
\caption{Proposed incremental fusion at penultimate layer for multimodal beamforming.}
\label{fig_fusion_incre}
\vspace{-5mm}
\end{figure}

\subsubsection{Incremental Fusion}
For incremental fusion network, we first sort unimodal networks according to their performance. We concatenate the penultimate layer embeddings of the two best unimodal networks. In this step, we freeze the best unimodal model and only retrain the second best unimodal model and the fusion layers. This forces the second-best unimodal model and the fusion layer to learn different information as compared to the best unimodal model. Similarly, when incorporating the third-best unimodal model, \textcolor{black}{we only retrain the third and fusion layers}. A conceptual overview of the incremental fusion framework is illustrated in Fig.~\ref{fig_fusion_incre}.
Following the same notations as aggregated fusion, the combined latent feature matrix $\mathbf{z}$ can be represented as:
\[
    \mathbf{z} = [\mathbf{z}_{\mathtt{IN}} ; \mathbf{z}_{\mathtt{L}}]\in \mathbb{R}^{d^{\ldr}+d^{\img}+d^{\crd}}, 
    \vspace{-2mm}
\]
where, $\mathbf{z}_{\mathtt{IN}}\in ([\mathbf{z}_{\ldr} ; \mathbf{z}_{\img}], [\mathbf{z}_{\ldr} ; \mathbf{z}_{\crd}], [\mathbf{z}_{\img} ; \mathbf{z}_{\crd}])$, and $\mathbf{z}_{\mathtt{L}}\in [\mathbf{z}_{\ldr}, \mathbf{z}_{\img}, \mathbf{z}_{\crd}]$. We denote the incremental fusion network as $f_{\theta^\IRF}^\IRF(.)$ as:
\[
 \scores_{\IRF} = \sigma (f_{\theta^\IRF}^\IRF(\mathbf{z})),~~~~~f_{\theta^\IRF}^\IRF:\mathbb{R}^{d^{\ldr}+d^{\img}+d^{\crd}} \mapsto \mathbb{R}^{|\mathcal{B}|} 
\]
where $\scores_{\IRF}$ is the ultimate layer output of the incremental fusion network.
\subsection{Multi-level Deep Fusion}


So far, the proposed feature-based fusion networks exploit the correlation in latent embeddings of the unimodal networks. However, such type of fusion networks can further be exploited along with the unimodal networks for an improved prediction at the second level. The output of the ultimate layers of each unimodal or fusion network represent the sensitivity of that particular model in detecting each beam pair. Hence, we propose a multi-level deep fusion technique which involve unimodal networks along with the penultimate fusion network.

\begin{figure}[hbtp]
\centering
\resizebox{0.4\textwidth}{!}{\includegraphics[width=0.48\textwidth]{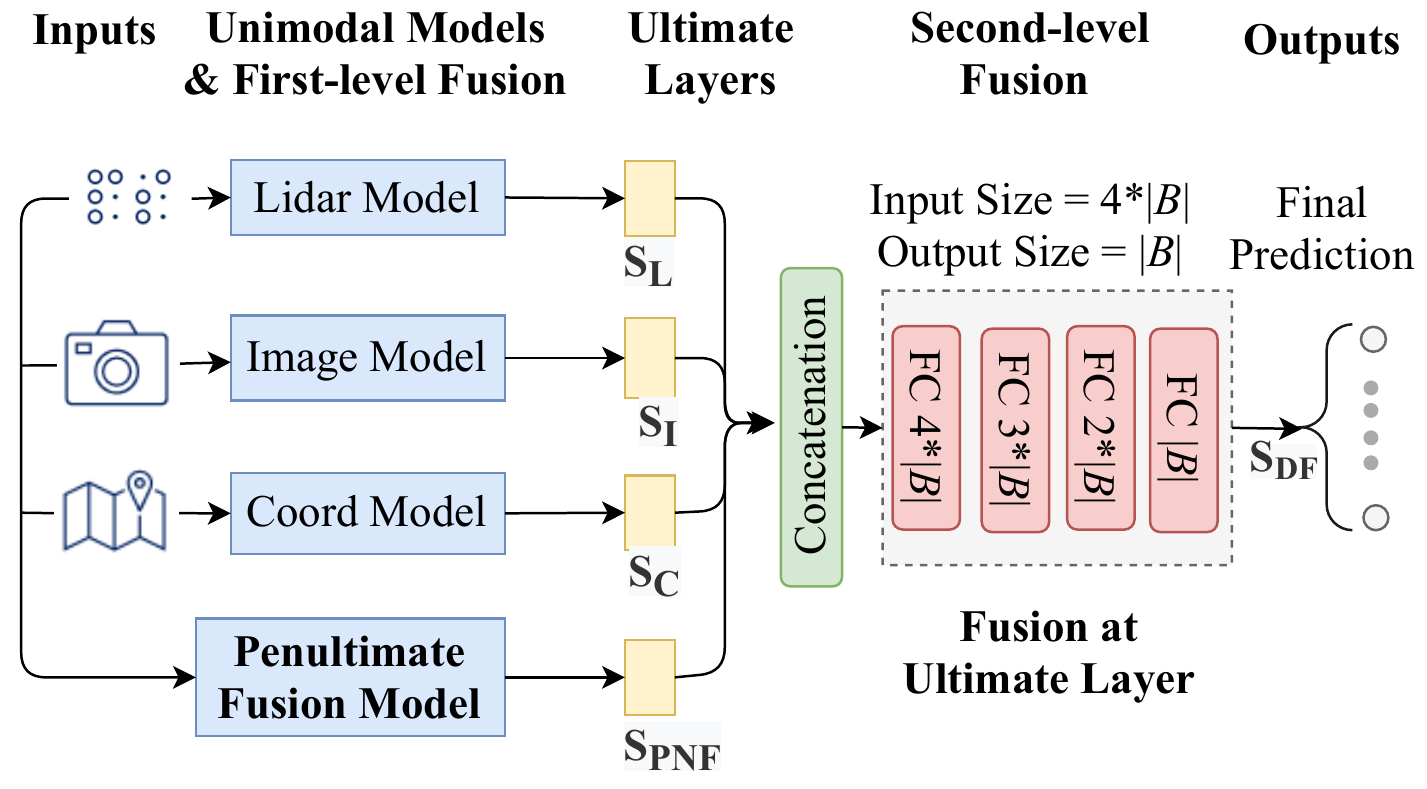}}
\caption{Proposed multi-level deep fusion framework at ultimate layers  for multimodal beamforming.}
\label{fig_fusion_deep}
\vspace{-5mm}
\end{figure}

The graphical representation of the deep fusion network at the second-level of fusion is shown in Fig.~\ref{fig_fusion_deep}. The deep fusion network on the output of ultimate layer will intelligently assign higher weights to the outputs of the more relevant models of first-level. We use 4 fully connected layers as the deep fusion architecture of second-level. The details of this network architecture is presented in Fig.~\ref{fig_fusion_deep}. The number of used filters in each level is flexible to be fine-tuned with the \textcolor{black}{available type of} dataset. 

The deep fusion at the second level is defined using the ultimate layers of unimodal and penultimate fusion networks: $\scores_{\ldr}$, $\scores_{\img}$, $\scores_{\crd}$, and $\scores_{\mathtt{PNF}}$ $\in \mathbb{R}^{|\mathcal{B}|}$, where $\scores_{\mathtt{PNF}} \in (\scores_{\AGF}, \scores_{\IRF})$.
In this case, the representation of combined matrix $\scores_{\DF}$ is:
\[
    \scores_{\DF} = [\scores_{\ldr}; \scores_{\img}; \scores_{\crd}; \scores_{\mathtt{PNF}}]\in \mathbb{R}^{4\times|\mathcal{B}|}.
\]
We denote the multi-level deep fusion network as $f_{\theta^\DF}^\DF(.)$. The softmax activation function ($\sigma$) is used to predict the optimality of each vehicle as:
\[
 \scores_{\DF} = \sigma (f_{\theta^\DF}^\DF(\mathbf{z})),~~~~~f_{\theta^\DF}^\DF:\mathbb{R}^{4\times|\mathcal{B}|} \mapsto \mathbb{R}^{|\mathcal{B}|} 
\]
where $\scores_{\DF}$ is the ultimate layer output after the second-level of fusion.

\subsection{Preliminary Results}
\begin{figure}[t!]
    \centering
     \resizebox{0.35\textwidth}{!}{\begin{tikzpicture}
      \begin{axis}[
      width  = 0.48*\textwidth,
      height = 6cm,
      major x tick style = transparent,
      ybar=2*\pgflinewidth,
      bar width=5pt,
      ymajorgrids = true,
      yminorgrids = true,
      minor tick num={2},
      minor y tick style={line width=2pt},
      ymin=0,
      ymax = 100,
      xmin =0,
      xmax = 6,
      xtick={1,2, 3, 4, 5,6},
        xticklabels={Top-1, , Top-5, ,Top-10, },
        x label style={at={(axis description cs:0.5,-0.1)},anchor=north},
      x tick label style={rotate=0, align=center,text width=1.2cm},
      ylabel= Accuracy (\%),
      legend cell align=left,
    legend style ={at={(0.5,-0.15)}, anchor=north, legend columns=-1},
  ]
      \addplot[style={fill=red!10!white},error bars/.cd, y dir=both, y explicit, error bar style={line width=1pt, blue}, error mark options={
      rotate=90,
      blue,
      mark size=2pt,
      line width=1pt
    }]
          coordinates {
          (1, 12.32) 
          (3, 55.61) 
          (5, 77.93)};
         
          \addplot[style={fill=cyan},error bars/.cd, y dir=both, y explicit, error bar style={line width=1pt, blue}, error mark options={
      rotate=90,
      blue,
      mark size=2pt,
      line width=1pt
    }]
          coordinates {
          (1, 12.39) 
          (3, 55.38) 
          (5, 71.65)};
    \addplot[style={fill=purple},error bars/.cd, y dir=both, y explicit, error bar style={line width=1pt, blue}, error mark options={
      rotate=90,
      blue,
      mark size=2pt,
      line width=1pt
    }]
          coordinates {
          (1, 46.23) 
          (3, 82.43) 
          (5, 89.95)};
    \addplot[style={fill=green!70!black},error bars/.cd, y dir=both, y explicit, error bar style={line width=1pt, blue}, error mark options={
      rotate=90,
      blue,
      mark size=2pt,
      line width=1pt
    }]
          coordinates {
          (1, 56.22) 
          (3, 85.53) 
          (5, 91.11)};

    \legend{Coordinate, Image, LiDAR, Aggregated Fusion}
  \end{axis}
  \end{tikzpicture}}
    \caption{Performance of different unimodal and aggregated fusion network on multimodal Raymobtime dataset~\cite{Klautau_2018}. We report the top-$K$ ($K$=1, 5, 10) accuracy of the predicted labels and true labels. Fusion significantly improves the top-1 accuracy, whereas the difference gets lesser while evaluating top-10.}
    \label{fig:prelm_results_unimodal}
    \vspace{-5mm}
\end{figure}
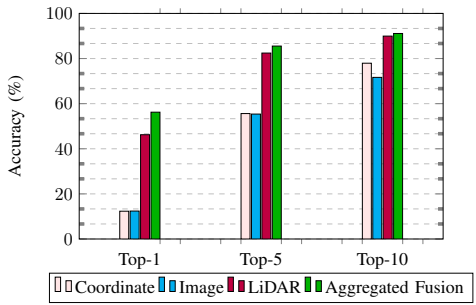

The preliminary results of predicting top-1, top-5, and top-10 best beams on the Raymobtime dataset~\cite{Klautau_2018} (details are in Sec.~\ref{subsec:raymobtime_dataset}) are presented in Fig.~\ref{fig:prelm_results_unimodal}. The initial results show the effectiveness of aggregated penultimate fusion over individual modalities. \textcolor{black}{LiDAR performs better than the other two sensors but fusion proves to be more effective instead of an LiDAR-only approach.}  The difference in performance of fusion and best performing modality decreases with  prediction of more beam pairs. Hence, to strategically decrease the search space for faster beam selection, fusion is more effective than any standalone, single modality. Implementation of the proposed incremental and multi-level fusion techniques on the same dataset are fertiled areas of further exploration by the research community.

\subsection{Future Possibilities: from Theory to Practice}
\begin{figure}[t!]
\centering
\includegraphics[width=0.48\textwidth, trim=1cm 4cm 8cm 0cm, clip]{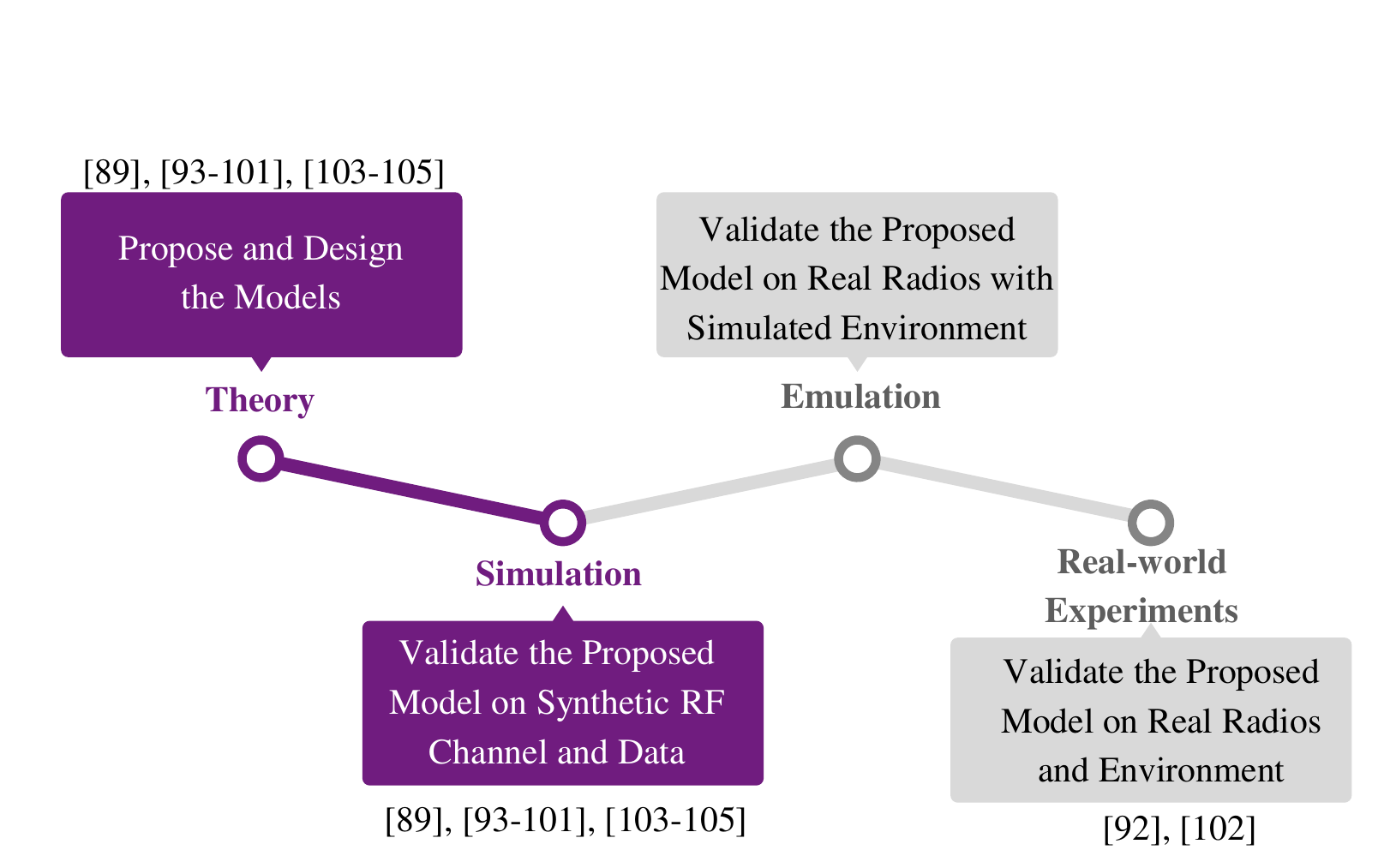}
\caption{Future possibilities: from theory to practice (literature is referred from Tab.~\ref{tab:survey_non_RF}). The multimodal beamforming can be shaped to NextG compatibility by further emulation and real world validation on the recent 5G standards.}
\label{fig_future_possibilities}
\vspace{-5mm}
\end{figure}

The pathway leading from theoretical conceptualization to real-world application is challenging given the four key  steps \textcolor{black}{involved}: (i) numerical analysis of theory; (ii) simulation on a system model; (iii) emulation of the system model using realistic setup; and (iv) validating the theory and system model with real-world testbed experiments. A typical ``path to reality" for wireless communication research is illustrated in Fig.~\ref{fig_future_possibilities}.
The state-of-the-art literature on beamforming using non-RF data is mainly confined within the first two stages, with very few efforts on real-world experiments. The most popular simulation dataset for numerical analysis and simulation is the multimodal Raymobtime dataset. However, it has many limitations: (i) it considers the simplistic mmWave geometric channel model, whereas real-world mmWave channels are more diverse in terms of propagation characteristics ; (ii) the average speed of the vehicles are limited to only $\sim$18 mph; (iii) the sampling rate of the sensors are very low (one sample in 30 seconds), which is not realistic for fast moving vehicles; and (iv) no pedestrian or vegetation are simulated in the environment.  

As of the {\em third stage}, one possible path forward for designing {\em emulation scenarios} with realistic 3D modeling may use the NVIDIA Omniverse~\cite{omniverse} platform to generate multimodal data with ground truth of RF ray tracing. However, the NVIDIA ray tracing engine Optix is optimized for ray tracing for photon atoms in visible light, which works at $>400$ THz frequency. Moreover, it does not support the diffraction property of RF waves. Hence, we need to explore different  integration scenarios for realistic emulation. The mmWave propagation models can be first generated via the RemCom wireless Insite software and then the NVIDIA Omniverse platform may provide the rest of the sensor data captures, motion and visual environment emulation. 

Another scope for  \textcolor{black}{innovation} is to collect {\em real-world multimodal data} comprising of LiDAR, camera image and GPS sensor data from an autonomous vehicle. These outdoor experiments can be undertaken in 60 GHz IEEE-V band that allow license-free use. A vehicle mounted setup of mmWave radios with RSU can be envisioned to collect multimodal sensor data from: (i) different locations in a city canyon region and (ii) in residential neighborhood region. The dataset should be inclusive of diversified wireless environments by considering different types of: (i) road curvature; (ii) pedestrian traffic; (iii) vehicular traffic from 5 mph to 45 mph; (iv) vegetation; (iv) weather condition. Such dataset can become the {\em defacto standard} for other researchers to comparatively benchmark and validate similar approaches for multimodal beamforming. 

Next, we discuss possible application areas of using multimodal beamforming in AI-enabled technologies for NextG networks. 


\section{Multimodal Beamforming: Applications}
\label{sec:applications}
In this section, we suggest applications where the benefits of beamforming with non-RF modalities are highly relevant. 

\subsection{Visual IoT} \label{subsec:visualIoT}
Current IoT developments rarely include visual data, even though powerful camera sensors have been developed over the years. The state-of-the-art cameras use dynamic, neuromorphic sensors that mimic the eye processes of mammals and can record billions of pixels per second. The key components of visual IoT produce large amounts of video data~\cite{visualIoT}. This necessitates rapid communication between visual IoT sensor nodes (edge devices), gateways and visual cloud servers~\cite{visualIoT2}. Communication between IoT sensors and the rest of the system becomes challenging when we consider the volume of visual data. Taking surveillance cameras (closed-circuit television) as an example, there are projections of 1 billion cameras being installed globally by 2025 - if we assume 12 hour recordings each day, we would produce 1250 petabytes of data ~\cite{IoB2018}. If this volume of data only comes from surveillance cameras, we can imagine the speed of data streams needed to communicate visual IoT sensor data to gateways and servers. 

Our topic of interest, multimodal beamforming with non-RF data, can offer a solution for the need of such ultra-fast communication between visual IoT sensors with high data rates. The ability to use non-RF data and leverage AI-based algorithms, to fuse the different data modalities, as part of the beamforming algorithms can revolutionize visual IoT and aid in implementation of smart cities. By enabling fast communication within networks of high-speed, high-resolution cameras with other modalities such as GPS or radar, multimodal beamforming can contribute to applications such as traffic management, emergency monitoring preparedness, air quality management and smart parking. Visual IoT sensors deployed for smart city applications include unmanned vehicles (UAVs and UGVs) traffic cameras, smart phone cameras and more. Non-RF multimodal beamforming enables those sensors to communicate in networks that provide city-wide coverage with low latency.

\subsection{V2X Architecture}
The Vehicle-to-Everything (V2X) market is estimated to be \$689 million in 2020 and projected to reach \$12,859 million by 2028~\cite{markets_2020}. V2X will enable  communication among vehicles as well as between vehicles and networks, infrastructure and pedestrians, aiming to improve traffic efficiency, road safety and individual vehicle energy efficiency \cite{v2x}. V2X connectivity is also essential for the advancement of autonomous driving. Traffic efficiency improves by monitoring congested areas and providing alternative routes, while maintaining road safety by monitoring speed and identifying risky drivers. At the same time, V2X networks can improve energy efficiency by making vehicles more intelligent, choosing journeys with lower carbon emissions.

\begin{figure}[t!]
\centering
\includegraphics[width=0.42\textwidth]{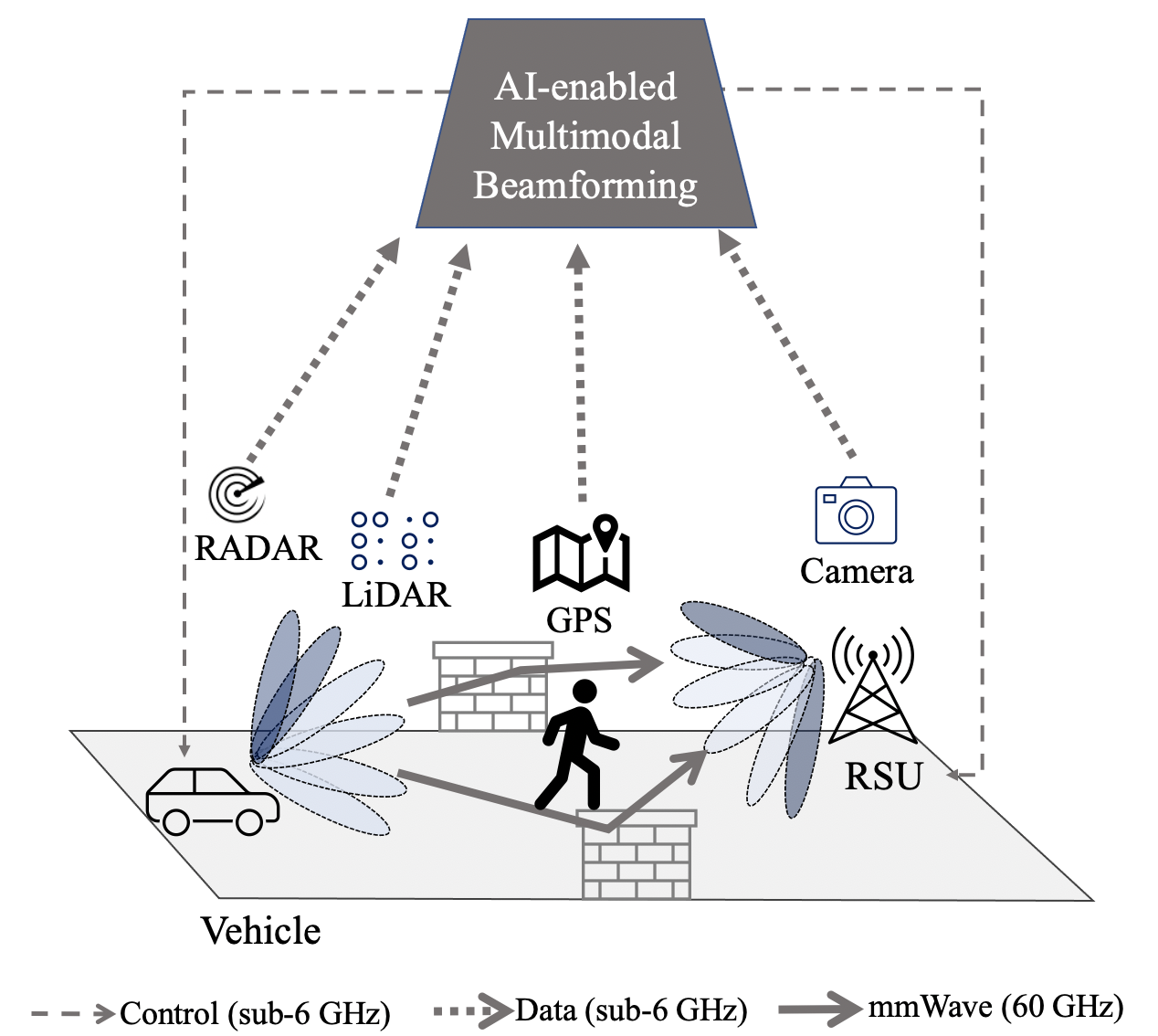}
\caption{Pedestrian detection using multimodal beamforming in V2X environment. Different sensor information (LiDAR, RADAR, Camera, GPS) are leveraged to perform the beamforming between vehicle and road side unit (RSU), as well as pedestrian detection between them.}
\label{fig_v2xapplication}
\vspace{-5mm}
\end{figure}


Different beamforming techniques has direct impact on the performance of 5G-V2X networks~\cite{v2x2}. 
In~\cite{Lee_2020}, Lee {\em et al.} presents an object detection algorithm by fusing visual and LiDAR data to form 3D images of the vehicle surroundings. Combining these two concepts, we envision the application area of using multimodal beamforming for V2X architecture will extend from fast and reliable communication to object detection in urban scenarios. As mentioned in Sec.~\ref{subsec:dataAcq_and_pre}, availability of different types of sensor data forms the backbone of V2X communication. The multimodal beamforming using these sensors can be leveraged to provide low latency V2X communication as well as knowledge of selected beams at a specific position can be leveraged to detect objects or pedestrian using AI-enabled algorithms. An example use-case of pedestrian detection via beamforming is depicted in Fig.~\ref{fig_v2xapplication}. 


\subsection{UAV Communication}
Unmanned aerial vehicles (UAVs) are used extensively in military, scientific and civil applications. They can be used for capturing data, monitoring non-accessible areas and developing high-throughput wireless communication infrastructure. Networks of UAVs, known as flying ad-hoc networks (FANETs) have sparked great interest in academia, industry and government due to their flexibility, low cost and wide range of applications: disaster management, relaying networks, agricultural processes and many more \cite{FANET}. For all those applications, high-speed low-latency wireless communication is essential between UAVs as well as from UAVs to ground entities (UAV-Ground). 

Images captured by flying UAVs may need to be distributed to ground nodes, while data from the ground terminals is required by the UAVs for channel allocation and routing  \cite{Tian2021ApplyingChallenges}. 
\textcolor{black}{Distributed beamforming is an important enabler for leveraging high throughput and long range communications through flying UAVs, given their high probability of LoS links due to their altitude. Drawbacks in these scenarios such as inaccurate GPS signals, unpredictable UAV hovering, etc., create the need for accurate transmission synchronization between multiple UAVs through external sensor data input~\cite{alemdar2021rfclock}, in order to realize a practical distributed beamforming implementation for multi-UAV to ground~\cite{mohanti2019airbeam} and UAV-UAV communications.} 
\cite{mmwave60} and \cite{mmwaveground} highlight the use of mmWave links for
UAV-UAV and UAV-ground communication. Ultra-fast UAV communication is essential for  wireless infrastructure drones (WIDs). To improve the need for faster and more reliable communication in both the above cases, beamforming in mmWave can be combined with AI-enabled techniques. With the introduction of camera images and other non-RF multimodal data, such as GPS, beamforming in UAV communication can be enhanced with multimodal beamforming to provide higher throughput, robustness, coverage and delay metrics. 
\vspace{-0.1 in}
\subsection{eXtended Reality (XR)}
The FCC plans to release up to 3 Thz spectrum that will accelerate future standardization efforts beyond 5G~\cite{horwitz_2019}. By achieving extreme data rates and high device capacity in NextG networks, the requirements for augmented reality (AR), virtual Reality (VR), and mixed reality (MR) (comprehensively extended reality (XR)) applications can be met \cite{6GAR}.  Terahertz~(THz) frequencies may support the bandwidth for wireless cognition - the real-time wireless transfer of human brain computations. Such high-speed communication increases the quality of physical experience, an important factor for AR/VR/MR/XR applications.

At the THz frequencies, the limited transmission distance once again requires directional antennas and beamforming with a higher number of antenna elements. Multimodal beamforming with AI-enabled algorithms at the THz/sub-THz band can be used in XR systems to ensure fast beam alignment to meet the real-time, high-speed data communication requirements.

\subsection{Multi-agent Robotics Applications}
Autonomous agents are increasingly used in a variety of applications like mining \cite{marshall2016robotics}, agriculture \cite{bonadies2016survey}, military \cite{young2016survey}, aerospace \cite{pedersen2003survey} and medicine \cite{liu2018multiagent} to name a few. Many system entities need to collectively coordinate with each other to make decisions online and collaboratively in this paradigm. Some examples of multi-agent robots are simultaneous localization and mapping (SLAM) \cite{brink2020maplets, bhutta2020loop, krinkin2017modern}, warehouse robotics \cite{chung2019multiagent, poudel2013coordinating, fujita2020important}, surgical robotics \cite{bodner2004first, rao2018robotic}, autonomous driving \cite{shalev2016safe, yu2019distributed, palanisamy2020multi}, agricultural robotics \cite{arguenon2006multi, afrin2021resource, davoodi2021graph} etc. In these   applications, each of the agents in a multi-agent system may be equipped with sensors like LiDAR, RGB and IR cameras, GPS receiver etc., which enable them to function autonomously. Many applications rely on agents being able to communicate locally with other agents. For such applications and for real-time collaboration, high-speed communication is needed for sharing sensor information, decisions and actions. To support such large data-rate requirements,  industries with highly automated process flows are pursuing high bandwidth communication links, including access to mmWave bands \cite{lu2020see}. Multimodal beamforming using non-RF data can be an interesting approach to facilitate the faster communication between such autonomous entities using the integrated sensors within them. 


\section{Emerging Research Frontiers}
\label{sec:emerging_trend}

In this section, we present selected research frontiers, where the AI-enabled multimodal beamforming techniques can make transformative difference.

\vspace*{-0.1in}
\subsection{Terahertz Communication}
Early works that prove the feasibility of  exploiting the THz frequency bands (0.3 THz to 10 THz) point towards an upcoming paradigm shift in the way wireless spectrum will be used. THz-band links bridge the gap between radio and optical frequency ranges, which may be game-changing for nextG wireless networks~\cite{Elayan_2018} by enabling transfer rates of 10Gb/s ~\cite{Song_2011}. However,  highly directional and fine-grained beams in the phased array antennas, which are essential to support the THz communication, come with their own challenges. Additionally, the beam search space increases with the increasing frequencies. Hence, there is urgent need to exploit  out-of-the-box approaches, such as AI-enabled CSI estimation techniques, to decouple the number of antenna elements from the beamforming time overhead~\cite{Belgiovine2021}. 
We believe the idea of multimodal beamforming can also be extended to THz communication to reduce the exploding search space of antenna codebook elements by leveraging the environmental multimodal data.


\vspace*{-0.1in}
\subsection{Virtual Presence}
Since the start of the COVID-19 pandemic, we have quickly transitioned to using virtual communications platforms to aid in  wellness and safety. However,  platforms like Zoom, Teams can only do so much with respect to quality of user experience. Most of these platforms are still limited by the on-screen presence. This is where the recent development of AR/MR/VR/XR can make a difference by opening up the possibility of transforming {\em on-screen presence} to a {\em virtual presence}. The concept of holographic representation can  emulate physical presence for meeting, gaming, or collaborating with others. Such virtual presence will supports mobility  while giving a group presentation or multi-player gaming. AR/MR/VR/XR technologies will require multi-Gbps data-rates that may  saturate a sub-mmWave band within  seconds. Even the still-evolving  5G standard is not capable of supporting these data transfer rates. The standardization of ultra-fast beamforming in mmWave communication is integral for NextG standards~\cite{Zhang_2020}. 
The concept of using multimodal non-RF data in such applications is  promising in this regard. The  rich properties of XR or holographic images can be exploited for situational awareness to aid in the beamforming in high frequencies, where the codebook search space is generally too large to compute optimally in real time via exhaustive searching~\cite{Lin_2017}.

\vspace*{-0.1in}
\subsection{Hybrid Beamforming}
mMIMO communications in hybrid transceivers is realized by a combination of high dimensional analog phase shifters and power amplifiers with lower-dimensional digital signal processing units~\cite{Ahmed_2018}. For fully connected hybrid transceivers, the situational states through the non-RF modalities can be leveraged to select multiple phase shifters (multi-label prediction), which can be inferred to derive the best RF chains and aid in even-faster beamforming. Multimodal beamforming can be applied per RF chain to select best phase shifter, and this will enable the parallel inference of all the RF chains at the same time. Hence, the use of multimodal data has huge potential for improving the emerging hybrid beamforming technique it will allow seamless scaling to make it suitable  for NextG networks.

\vspace*{-0.1in}
\subsection{Multiple User Massive MIMO (MU-MIMO)}
Massive MIMO links for multiple concurrent users will soon become part of the 5G standard~\cite{Lu_2014}. The open challenge of making interference-free beam formation with multiple users with available RF chains is the main roadblock in  MU-MIMO. Another challenge is to guarantee scalable, real-time signal processing in large MU-MIMO systems~\cite{Qing_2013}. Similar to the other trends, the situational state information from different non-RF modalities (such as image, infrared, LiDAR) can be used to build trained models to address these challenges, while providing scalability with number of antennas and users.


\section{Conlcusions}
\label{sec:conclusions}
This paper provides a comprehensive survey of using AI-enabled beamforming techniques for  out-of-band and multimodal non-RF data RF for mmWave-band operation in NextG networks. While the previous surveys on beamforming ~\cite{Molisch_2017}, \cite{Kutty_2016}, \cite{Ahmed_2018} are focused more on analyzing and using mmWave channel characteristics, or channel state information for beamforming in massive MIMO leveraging the complicated hybrid beamforming process; our survey reviews rercent trends in the literature that adopt an out-of-box approach for solving the same problem. We discuss the state-of-the-art in research trends, application areas, and open challenges of this exciting  and emerging paradigm of multimodal sensor data-enabled beamforming. We also present novel AI-enabled fusion techniques which prove\st{s} the effectiveness of exploiting multimodal data for non-RF data based beamforming. We identify open research challenges to motivate future researches and well as indicate the potential transformative impact of this area on different wireless applications.


\section*{Acknowledgment}
This material is based upon the research sponsored by the Nvidia Inc.
\vspace*{-0.05in}


\bibliographystyle{IEEEtran}
\bibliography{references, references_mendeley}
\end{document}